\documentclass[aps,prd,12pt,citeautoscript,superscriptaddress,showpacs]{revtex4-1}
\usepackage{xcolor}
\usepackage[colorlinks=true, pdfstartview=FitV,linkcolor=blue, citecolor=blue, urlcolor=blue]{hyperref}

\newcommand{\be}{\begin{equation}}
\newcommand{\ee}{\end{equation}}
\newcommand{\bea}{\begin{eqnarray}}
\newcommand{\eea}{\end{eqnarray}}

\newcommand{\bs}{\boldsymbol}
\newcommand{\h}{\gamma}
\newcommand{\hh}{h}
\newcommand{\nuu}{\nu}
\newcommand{\pii}{p}

\usepackage{psfrag}
\usepackage{epsfig}
\usepackage{latexsym,array,theorem,mathrsfs,subfigure,bm,float}
\usepackage{amsmath}
\usepackage{amsfonts}
\usepackage{amssymb}
\usepackage{bbm}
\usepackage{color}

\renewcommand{\theequation}{\arabic{section}.\arabic{equation}}

\begin{document}
\title{Energy in ghost-free massive gravity theory}

\author{Mikhail~S.~Volkov}
\email{\tt volkov@lmpt.univ-tours.fr}
\affiliation{
Laboratoire de Math\'{e}matiques et Physique Th\'{e}orique CNRS-UMR 7350, \\ 
Universit\'{e} de Tours, Parc de Grandmont, 37200 Tours, FRANCE}
\affiliation{
Department of General Relativity and Gravitation, Institute of Physics,\\
Kazan Federal University, Kremlevskaya str.18, 420008 Kazan, RUSSIA
}


\begin{abstract} 
The detailed calculations of the energy  
in the ghost-free massive gravity theory is presented.  
The energy is defined in the standard way within the canonical 
approach, but to evaluate it requires resolving  
the Hamiltonian constraints, which are   
known, in general, only implicitly. 
Fortunately, the constraints can be explicitly obtained
and resolved in the spherically symmetric sector, which allows one to 
evaluate the energy. It turns out that the energy is positive 
for globally regular and asymptotically flat fields
 constituting 
the ``physical sector" of the theory.  
In other cases the energy can be 
negative and even unbounded from below, 
which suggests that the theory could be still plagued with ghost instaility.
However, 
a detailed inspection reveals that 
the corresponding solutions of the constraints are either not globally regular or not 
asymptotically flat. Such solutions cannot describe initial data 
triggering ghost instability of the physical sector.
This allows one to conjecture that the physical sector could actually be protected from 
the instability by a potential barrier separating it from negative energy states.

\end{abstract}


\maketitle

\section{Introduction}

The idea that gravitons could have a tiny mass, 
which would explain the current cosmic acceleration 
\cite{1538-3881-116-3-1009,*0004-637X-517-2-565}, 
has attracted a lot of interest  after the 
discovery of the special massive gravity theory 
by de Rham, Gabadadze, and Tolley (dRGT) 
\cite{deRham:2010kj} 
(see \cite{Hinterbichler:2011tt,*deRham:2014zqa}
for a review).
Before this discovery it had been known that the massive gravity theory
generically  
had six  propagating degrees of freedom (DOF). Five of them could be 
associated with the polarizations of the massive graviton, while the sixth 
one,  usually called Boulware-Deser (BD) ghost, is non-physical, because 
it has a negative kinetic energy and  renders the whole theory unstable. 
The speciality of the dRGT theory is that it  
contains two Hamiltonian constraints 
 which eliminate one of the six DOFs
\cite{Hassan:2011hr,*Hassan:2011ea,*Kluson:2012wf,*Comelli:2012vz,*Comelli:2013txa}.
Therefore, there remain just the right 
number of DOFs to describe massive gravitons, and
as the theory does not show non-physical features in special limits,
it is referred to as ghost-free.  

However, the fact that the theory has the correct number of DOFs does not 
yet guarantees that they are all physical and always behave correctly. It is possible that 
the DOF removed by the constraints is not exactly the BD ghost 
but its superposition with physical modes. Therefore,   
 it could be that the 
remaining five  DOFs are still 
contaminated with a remnant of the ghost, 
suppressed in some cases but present otherwise.  
Unfortunately, such a concern is supported by the observations of certain    
ghost-type features in the theory    
\cite{DeFelice:2012mx,*Fasiello:2013woa,*Chamseddine:2013lid}. 

A good way to see whether the theory is indeed 
ghost-free is to compute the energy since  if the energy is positive, 
the ghost is absent. 
The energy can be straightforwardly  defined within the standard canonical 
formalism of Arnowitt-Deser-Misner (ADM) \cite{ADM}. 
However, the problem is that to evaluate the energy 
requires resolving the constraints, 
which are known, in general, only implicitly.
For this reason 
the energy in the theory 
has never been computed.  
Therefore, our aim  
is to compute it in the spherically symmetric
sector (the s-sector), where the constraints can be obtained explicitly 
and, in some cases, resolved.  
The corresponding solutions can be viewed as initial data 
for the Cauchy problem.

It turns out that the energy is positive 
for globally regular and asymptotically flat  solutions of the 
constraint equations.  All of such solutions constitute the  
``physical sector" of the theory.  At the same time, there are also 
other solutions of the constraints for which the energy 
can be negative and even unbounded from below. In addition, for certain negative energy 
solutions the Fierz-Pauli (FP) mass becomes imaginary 
so that the gravitons effectively behave as 
tachyons. This reminds  of the recent finding of the superluminal waves 
in the theory \cite{Deser:2012qx,*Deser:2013eua,*Deser:2013qza,*Deser:2014hga}. 
At first glance, one can think that all of this indicates that the 
theory is still plagued with the ghost. 
However, a closer inspection reveals that the negative energy solutions
of the constraint equations are always either not globally regular or not 
asymptotically flat. Such solutions are unacceptable as initial data 
for perturbations around the flat space, hence  they 
cannot affect the physical sector.

{
This suggests that the physical sector could actually be protected from 
ghost instability by a potential barrier separating it from sectors 
containing negative energy states. 
If true, this would mean that the physical sector should be 
protected also from the tachyons, as they have negative energies.
Moreover, it would follow that  the physical sector could be 
protected from the superluminal waves as well,
as they presumably coexist with the tachyons. 
Therefore, it is possible that the negative energies 
and other seemingly non-physical features do not actually invalidate the whole
theory since they do not affect 
the physical sector.} 
At the same time, one should emphasise that all of these arguments 
can only be viewed as a conjecture currently supported 
only by evidence found in the $s$-sector. 
The main body of the paper below is devoted to the detailed calculations,
whose results could be interpreted as indicated above.

The rest of the paper is organized as follows. After a brief description
of the massive gravity theory in Section \ref{Sec1}, its Hamiltonian formulation 
is discussed in Section \ref{Sec2}, focusing  on the comparison 
of the generic massive gravity, the FP theory, and the dRGT theory. 
Section \ref{Sec3} describes the reduction to the 
s-sector and computation of the constraints, whose 
weak field limit   is described in Section \ref{Sec4}. 
The next two Sections describe what happens away from the weak field limit.
Section \ref{Sec5} considers the kinetic energy sector where the metric is fixed 
but the momenta can vary. The solutions of the 
constraints then split into two disjoint branches, one with positive and one
with negative energies. The negative energies can be arbitrarily large, 
however,  the corresponding solutions of the constraints are singular. 

Section \ref{Sec6} considers the 
potential energy sector where the momenta vanish but the metric  
can vary. In this sector, too, there are two branches of solutions of 
the constraints: the positive energy branch containing the flat space, 
and the ``tachyon branch'' containing a special 
solution with a  constant and negative energy density.   
In addition, there are asymptotically flat 
``tachyon bubbles'' with negative energies which 
interpolate between the two branches. 
Their existence suggests at first that the flat space could decay into bubbles, but 
a closer inspection reveals that the corresponding initial data are singular
and cannot describe the decay process. Section \ref{Sec7} contains concluding remarks,
and many  technical details are given in the five  Appendices. 

The unitary gauge for the reference metric is used all through the text.  
A brief summary of the results presented below can be found in 
Ref.\cite{Volkov:2014qca}

\section{Massive gravity \label{Sec1}}	
\setcounter{equation}{0} 
The theory is defined 
by the action 
\bea                                      \label{1}
&&S
=M^2_{\rm Pl}\int \, \sqrt{-{ g}}\left(\frac{1}{2}\,R-{m^2}\, {\cal U}\right)\,d^4x
\equiv M^2_{\rm Pl}\int {\cal L}\, d^4 x\,.
\,
\eea
Here the potential ${\cal U}$ is a scalar function of 
$H^\mu_{~\nu}=\delta^\mu_\nu
-g^{\mu\alpha}f_{\alpha\nu}$ of the form 
\be                            \label{PF}
{\cal U}= \frac18\,
(H^{\mu}_{~\nu}H^{\nu}_{~\mu}-(H^\mu_{~\mu})^2)
+\ldots ,
\ee 
where $f_{\mu\nu}$ is the flat reference metric, and 
the dots denote all possible 
higher order scalars made of $H^\mu_{~\nu}$. Such a form of the 
potential insures that in the weak field limit 
the linear FP theory \cite{Fierz:1939ix} 
of massive gravitons with 5 polarizations is recovered. 
However, away from the weak field limit and for the generic potential \eqref{PF}
the theory propagates 5+1 DOFs, the extra DOF
being the BD ghost \cite{Boulware:1973my}. 
At the same time, there is a unique choice of the higher order terms 
in \eqref{PF} for which, even at the non-linear level, the theory 
propagates only 5 DOFs.  
This special choice determines the dRGT 
theory \cite{deRham:2010kj}, in which case 
\be                             \label{2}
{\cal U}=\sum_{k=0}^4 b_k\,{\cal U}_k(\gamma),
\ee
where $b_k$ are parameters and  
\bea                        \label{4}
{\cal U}_0(\bs{\gamma})&=&1,~~~~~
{\cal U}_1(\bs{\gamma})=
\sum_{A}\lambda_A=[\bs{\gamma}],\nonumber \\
{\cal U}_2(\bs{\gamma})&=&
\sum_{A<B}\lambda_A\lambda_B 
=\frac{1}{2!}([\bs{\gamma}]^2-[\bs{\gamma}^2]),\nonumber \\
{\cal U}_3(\bs{\gamma})&=&
\sum_{A<B<C}\lambda_A\lambda_B\lambda_C
=
\frac{1}{3!}([\bs{\gamma}]^3-3[\bs{\gamma}][\bs{\gamma}^2]+2[\bs{\gamma}^3]),\nonumber \\
{\cal U}_4(\bs{\gamma})&=&
\lambda_0\lambda_1\lambda_2\lambda_3
=
\frac{1}{4!}([\bs{\gamma}]^4-6[\bs{\gamma}]^2[\bs{\gamma}^2]
+8[\bs{\gamma}][\bs{\gamma}^3]+3[\bs{\gamma}^2]^2
-6[\bs{\gamma}^4])\,. 
\eea
Here $\lambda_A$ are eigenvalues of 
$
\bs{\gamma}^\mu_{~\nu}=
\sqrt{{{g}}^{\mu\alpha}{{f}}_{\alpha\nu}}
$, 
with
 the 
square root
understood in the sense that  
\be                     \label{gg}
\bs{\gamma}^\mu_{~\alpha}\bs{\gamma}^\alpha_{~\nu}
={{g}}^{\mu\alpha}{{f}}_{\alpha\nu}.
\ee 
Using the hat to denote matrices one has  
$[\bs{\gamma}]\equiv {\rm tr}(\hat{\bs{\gamma}})= \bs{\gamma}^\mu_{~\mu}$, 
$[\bs{\gamma}^k]\equiv {\rm tr}(\hat{\bs{\gamma}}^k)= (\bs{\gamma}^k)^\mu_{~\mu}$. 
If the bare cosmological term
is absent, the flat space is a solution of the theory,
and $m$ in \eqref{1} is the FP mass of the gravitons 
in the weak field limit, 
then the coefficients $b_k$ in \eqref{2} can be expressed 
in terms of two arbitrary parameters, usually called $c_3$ and $c_4$, 
as
\be                    \label{bbb}
b_0=4c_3+c_4-6,~~
b_1=3-3c_3-c_4,~~
b_2=2c_3+c_4-1,~~
b_3=-c_3-c_4,~~ 
b_4=c_4.
\ee

\section{Hamiltonian formulation \label{Sec2}} 
\setcounter{equation}{0}
In order to pass to the Hamiltonian description of the theory \eqref{1},
one employs  the standard ADM decomposition 
of the spacetime metric $g_{\mu\nu}$ \cite{ADM}, 
\be                            \label{ADM}
ds_g^2=-N^2dt^2+\h_{ik}(dx^i+N^i dt)(dx^k+N^k dt),
\ee
($\h_{ik}$ is not to be confused with $\bs{\gamma}^\mu_{~\nu}$ in \eqref{gg}). 
The flat reference metric $f_{\mu\nu}$ is 
\be                            \label{ADM-f}
ds_f^2=\eta_{ab}\,\partial_\mu {\Phi}^a\partial_\nu {\Phi}^b dx^\mu dx^\nu \,, 
\ee
where $\eta_{ab}={\rm diag}[-1,1,1,1]$ and $\Phi^a(x^\mu)$ are fixed 
non-dynamical (in our approach) scalar fields (Stueckelberg scalars), 
whose choice determines the coordinate
system. 
Using these expressions, 
the Lagrangian density in \eqref{1} becomes 
\be
{\cal L} =\frac12\sqrt{\h}N\{K_{ik}K^{ik}-K^2+R^{(3)}\}
-m^2{\cal V}(N^\nu,\h_{ik})
+\mbox{total derivative}\,,
\ee
with ${\cal V}=\sqrt{\h}N{\cal U}$. 
Here $N^\mu=(N,N^k)$ are the lapse and shift functions, $K_{ik}$ is 
the second fundamental form of the hypersurface of constant time (see Eq.\eqref{A4}
in the Appendix \ref{ApA}),  
and 
$R^{(3)}$ 
is the Ricci scalar  
of $\h_{ik}$.
The indices are moved with $\h_{ik}$, and $K=K^i_i$. 
The Hamiltonian 
density is ${\cal H}=\pi^{ik}\dot{\h}_{ik}-{\cal L}$.
Explicitly,  
\be                           \label{Hm} 
{\cal H}
=N^\mu{\cal H}_\mu
+m^2{\cal V}\,,
\ee
where 
\be                          \label{Hm1}
{\cal H}_0=\frac{1}{\sqrt{\h}}\,
(2\pi^{ik}\pi_{ik}-(\pi^k_k)^2)-\frac12\,\sqrt{\h}R^{(3)},~~~
{\cal H}_k=-2\nabla^{(3)}_i \pi^i_k\,,
\ee
with the momenta conjugate to $h_{ik}$ 
\be                               \label{Hm2}
\pi^{ik}=\frac{\partial {\cal L}}{\partial\dot{\h}_{ik}}
=\frac12\,\sqrt{\h}\,(K^{ik}-K \h^{ik}).
\ee
The momenta conjugate to $N^\mu$
vanish, 
$
{\partial {\cal L}}/{\partial\dot{N}_{\mu}}=0,
$
so that $N^\mu$ are non-dynamical. Therefore,  
the phase space is spanned by 12 variables $(\pi^{ik},\h_{ik})$. 
Since the momenta conjugate to $N^\mu$ vanish, their time derivatives
should vanish as well. On the other hand, the  time derivatives of the 
momenta are obtained by varying  the Hamiltonian with respect to 
the conjugate to them variables. This requires that 
\be                    \label{HN}
\frac{\partial {\cal H}}{\partial N^\mu}=
{\cal H}_\mu(\pi^{ik},\h_{ik})
+{m^2}\frac{\partial {\cal V}
(N^\alpha,\h_{ik})}{\partial N^\mu}=0.
\ee
These conditions determine the number of propagating DOFs in the theory.

The energy is the Hamiltonian, 
\be
E=H=\int {\cal H}\,d^3x\,,
\ee
where the arguments of ${\cal H}$ should fulfill the conditions \eqref{HN}. 
For $m=0$ this expression for the energy should be 
augmented by the surface term needed to take into account the slow (Newtonian) 
asymptotic falloff of the fields when varying the Hamiltonian \cite{Regge:1974zd}. 
For $m\neq 0$ the falloff is exponential and no surface term is needed. 

It is instructive to consider particular cases.  

\subsection{General Relativity}

If $m=0$ then Eqs.\eqref{HN} reduce to
\be
{\cal H}_\mu(\pi^{ik},\h_{ik})=0,
\ee
which are four constraints for the phase space variables 
$(\pi^{ik},\h_{ik})$. 
These constraints  are first class, because their mutual Poisson brackets 
form an algebra, 
\be                                                \label{algebra}
\{ {\cal H}_\mu,{\cal H}_\nu \}_{\rm PB} =\sum_\alpha C^\alpha_{\mu\nu} {\cal H}_\alpha
\ee
(see \cite{Khoury:2011ay} for an explicit computation of the 
structure coefficients  $C^\alpha_{\mu\nu}$). 
First class constraints 
generate gauge symmetries, which allows one to 
impose in addition {four gauge conditions} on  $(\pi^{ik},\h_{ik})$ by fixing the gauge.
As a result, there remain
$
12-4-4=4
$
independent phase space variables;
they describe two graviton polarizations. 
The energy vanishes on the constraint surface,
$
{\cal H}=N^\mu{\cal H}_\mu=0 
$
(up to the surface term \cite{Regge:1974zd}). 

\subsection{Generic massive gravity}
If $m\neq 0$ then \eqref{HN}  
are not constraints but rather equations for the lapse and shifts, 
whose solution is 
$
N^\mu=N^\mu(\pi^{ik},\h_{ik} ).
$
Since there are no constraints, all twelve 
phase space variables $(\pi^{ik},\h_{ik})$
are independent and describe 
$6=5+1$ DOFs. 
These correspond to the five graviton 
polarizations plus one extra state.

Inserting $N^\mu(\pi^{ik},\h_{ik} )$ 
to ${\cal H}
=N^\mu{\cal H}_\mu
+m^2{\cal V}\,$
gives 
${\cal H}={\cal H}(\pi^{ik},\h_{ik})$, which 
turns out to be a non-positive definite function. 
In particular, ${\cal H}(\pi^{ik},\h_{ik})$   
can be made negative and arbitrarily large
by varying the momenta only, so that the 
kinetic energy is not positive definite 
\cite{Boulware:1973my}. Since the Hamiltonian 
is unbounded from below, the theory is unstable.  
This feature can be 
 attributed
to the extra DOF,
the BD ghost. One can  expect that if the 
ghost is eliminated in some way and only five DOFs remain, 
then the energy should be 
positive.

\subsection{Fierz-Pauli theory}

The analysis of the previous subsection goes differently 
in the linear FP theory, because constraints then arise. 
This theory can be obtained by expanding the Hamiltonian density \eqref{Hm}
around the flat space and keeping only the quadratic terms. 
Let us choose a static but  
not necessarily Lorentzian coordinate system, so that 
the flat f-metric reads
\be
ds_f^2=-dt^2+f_{ik}\,dx^idx^k\,,
\ee 
where $f_{ik}$ depend on $x^k$. 
The g-metric \eqref{ADM} is assumed to be close to the f-metric, so that  
\be
N=1+\nuu,~~~~N^k=\nuu^k,~~~~ 
\h_{ik}=f_{ik}+\hh_{ik}\,,
\ee 
where 
$\nuu$, $\nuu^k$, $\hh_{ik}$ and also 
the momenta $\pi^{ik}$ are small. 
 Let us expand ${\cal H}$ in \eqref{Hm},\eqref{Hm1} with respect to the small 
quantities.  
One has 
\be 
-\frac12\,\sqrt{\h}\,R^{(3)}=
\sqrt{f}\,(V_1+V_2)+\ldots ,
\ee
where the dots stand for higher order terms, while 
the first and second order terms are
\bea
V_1&=&\frac{1}{2}\,(\nabla^k\nabla_k h-\nabla^i\nabla^k h_{ik}),\\
V_2&=&\frac14\,h^{ik}\left(
-\frac12\,\nabla^s\nabla_s h_{ik}
+\frac12\,f_{ik}\nabla^s\nabla_s h
-\nabla_i\nabla_k h+\nabla_i\nabla^s h_{sk} 
\right).
\eea 
Here $\nabla_k$ is the covariant 
derivative with respect to $f_{ik}$, the indices 
are moved by $f_{ik}$, while $h=h^k_k$.   Components of 
the tensor $H^\mu_{~\nu}=\delta^\mu_\nu
-g^{\mu\alpha}f_{\alpha\nu}$ are
\be
H^0_0=2\nu+\ldots,~~~~~H^0_k=-\nu_k+\ldots,~~~~~
H^k_0=\nu^k+\ldots,~~~~~H^i_k=h^i_k+\ldots,
\ee
 so that the potential 
\eqref{PF} 
is
\be 
{\cal U}=\left.\left.\frac18\,
\right(h^i_k h^k_i-h^2-2\,\nu_k\,\nu^k-4\nu h\right)
+\ldots 
\ee
Inserting the above expressions to ${\cal H}$ in \eqref{Hm},
dropping the total derivative and keeping only the
quadratic terms, yields  
 the FP Hamiltonian density, 
\bea
{\cal H}_{\rm FP}&=&
\frac{1}{\sqrt{f}}\,(2\,\pi^i_k\pi^k_i
-(\pi^k_k)^2)
+\sqrt{f}\left(V_2+\frac{m^2}{8}(h^i_k h^k_i-h^2
-2\nu_k\nu^k )  \right)\nonumber \\
&-&\nabla_k\pi^k_s\,\nu^s
+\nu\sqrt{f}\left(V_1-\frac{m^2}{2}\,h
\right).
               \label{Hfp}
\eea
The crucial point is that the lapse 
$\nuu$ enters ${\cal H}_{\rm PF}$
linearly. Therefore, varying with respect to 
it gives a constraint,  
\be                        \label{CFP}
{\cal C}_{\rm FP}\equiv\frac{\partial{\cal H}_{\rm PF}}{\partial \nu}= 
\left.\left.
\frac12\,\sqrt{f}\right(
\nabla^k\nabla_k h-\nabla^i\nabla^k h_{ik}
-{m^2}\hh\right)=0\,.
\ee 
On the other hand, varying with respect to the shifts $\nuu_k$ 
gives equations with the solution
\be 
\nuu_k=-\frac{4}{{m^2}\sqrt{f}}\,
\nabla_m{\pi}^m_k\,.
\ee
Inserting this into ${\cal H}_{\rm FP}$
and dropping total derivatives yields 
\be                          \label{E-FP}
{\cal H}_{\rm FP}=
{\cal H}_{\rm FP}(\pi^{ik})
+{\cal H}_{\rm FP}(\hh_{ik})+\nuu\,{\cal C}_{\rm FP}\,,
\ee 
where 
\bea                     \label{Hfp1}
{\cal H}_{\rm FP}(\pi^{ik})&=&\frac{1}{\sqrt{f}}\left(2\,\pi^i_k\pi^k_i
-(\pi^k_k)^2+
\frac{4}{m^2}\,
\nabla_i{\pi}^i_k\nabla^j\pi_j^k\right), \\
{\cal H}_{\rm FP}(\hh_{ik})&=&\sqrt{f}\left(\frac18\,
\nabla^j h^i_k\nabla_j h^k_i
-\frac18\,\nabla_kh\nabla^k h
+\frac14\,\nabla_j\hh^j_k\,\nabla^k\hh \right.\nonumber \\
&-&\left.\frac14\,\nabla_j\hh^j_k\nabla^i \hh^k_i
+\frac{m^2}{8}\left(\hh^i_k\hh^k_i-\hh^2\right)\right).
  \nonumber                   
\eea
The ${\cal C}_{\rm FP}=0$ constraint should be preserved in time, therefore 
the Poisson bracket of ${\cal C}_{\rm FP}$ (see the Appendix \ref{ApE}) 
with 
${H}_{\rm FP}=\int {\cal H}_{\rm FP}\,d^3 x$ should vanish. This gives the 
secondary constraint,
\be                         \label{SFP}
{\cal S}_{\rm FP}\equiv \{{\cal C}_{\rm FP},
{H}_{\rm FP} \}_{\rm PB}= {m^2}\pi^k_k
+2\,\nabla^{i}\nabla^k{\pi}_{ik}=0\,.
\ee
The stability of this constraint
does not lead to new constraints but 
 yields an equation,
\be 
\{{\cal S}_{\rm FP},{H}_{\rm FP} \}= 
\frac{3}{4}\,m^4(h-\nuu)+\frac32\,m^2\partial^2_{kk} h+(\partial_{kk}^2)^2h=0,
\ee
which determines the lapse $\nu$.  
One has ${\{\cal C}_{\rm FP},
{\cal S}_{\rm FP}\}_{\rm PB}\neq 0$,
therefore the constraints are second class. 
Their existence 
implies that the number of independent phase space variables 
is $12-2=10$, so that there are five DOFs, which matches the number of   
polarizations of the massive graviton. 

Since the theory has  the right number 
of DOFs, one can expect the energy to be positive.
The positivity of the energy is in fact encoded in the FP theory 
by construction \cite{Fierz:1939ix},\cite{VanNieuwenhuizen:1973fi},
but one can also directly check that the energy is positive 
(see Appendix \ref{ApFP}).


\subsection{dRGT theory}

It turns out that for the potential \eqref{2}
the Hessian matrix 
$$
\frac{\partial^2{\cal V}(N^\alpha,h_{ik})}
{\partial N^\mu \partial N^\nu}
$$
has rank three \cite{deRham:2010kj}. For this reason the equations 
\eqref{HN}
\be                           \label{HN1}
{\cal H}_\mu(\pi^{ik},h_{ik})
+{m^2}\frac{\partial {\cal V}(N^\alpha,h_{ik})}{\partial N^\mu}=0~~
\ee
determine only the shifts, 
\be                                \label{Nk}
N^k=N^k(N,\pi^{ik},h_{ik} ),
\ee
whereas the lapse 
$N$ remains undetermined 
\cite{Hassan:2011ea,*Kluson:2012wf,*Comelli:2012vz}. 
Inserting $N^k$
into ${\cal H}=N^\mu{\cal H}_\mu+m^2{\cal V}$, 
the result has the structure 
$$
{\cal H}={\cal E}(\pi^{ik},h_{ik} )+N{\cal C}(\pi^{ik},h_{ik})\,.
$$
Varying this with respect to $N$ gives 
the  constraint 
\be                            \label{con-C}
{\cal C}(\pi^{ik},h_{ik})=0.
\ee 
Computing its Poisson brackets with 
${H}=\int {\cal H}\, d^3x$ 
gives the secondary constraint, 
\be                             \label{con-S}
{\cal S}(\pi^{ik},h_{ik})\equiv \{{\cal C},{H} \}_{\rm FP}=0\,,
\ee
while the condition  $\{{\cal S},{H} \}_{\rm FP}=0$
gives an equation for $N$. 
The two constraints 
 eliminate one DOF, hence there are only five propagating DOFs, 
as in the FP case, but this time at the fully non-linear level.

There remains to see if the energy is positive. 
 The energy is 
$$
H=\int {\cal E}(\pi^{ik},h_{ik} )\, d^3x\,,
$$
where $\pi^{ik},h_{ik}$ should fulfill the constraints \eqref{con-C}
and \eqref{con-S}. This latter condition renders computation of the energy 
extremely difficult since the constraints are non-linear partial 
differential equations which are hard to resolve. In addition, these 
equations are not known explicitly. The problem is that the 
equations \eqref{HN1} for the shifts $N^k$ are complicated and can be 
solved only in principle. This means that their solution 
exists, but its explicit form is not known,
unless for special values of the parameters $b_k$ \cite{Hassan:2011ea}. 
Therefore, neither the constraints nor the energy density are known
explicitly, which is why the energy in the theory has never been
computed.  For this reason we shall restrict ourselves to the spherically 
symmetric sector, where explicit expressions can be obtained.

\section{Spherical symmetry \label{Sec3}}
\setcounter{equation}{0}
Assuming spherical coordinates 
$x^\mu=(t,r,\vartheta,\varphi)$, one can parametrize 
the two metrics as 
\bea 
ds_g^2&=&-N^2dt^2+\frac{1}{\Delta^2}(dr+\beta\, dt)^2
+R^2 d\Omega^2\,,
\label{sphere} \\
ds_f^2&=&-dt^2+dr^2+r^2 d\Omega^2\,, \label{sphere-flat}
\eea 
where $N,\beta,\Delta$, and $R$ depend on $t$ and $r$; one has 
$d\Omega^2=
d\vartheta^2+\sin^2\vartheta d\varphi^2$. 
The dynamical variables can be chosen to be $\Delta,R$, 
with the conjugate
momenta (see the Appendix \ref{ApA})
\be
\pii_\Delta=\frac{\partial L}{\partial \dot{\Delta}}\,,
~~~~
\pii_R=\frac{\partial L}{\partial \dot{R}}\,.
\ee
The phase space is spanned by four  variables  
$(\Delta,R,\pii_\Delta,\pii_R)$, 
while $N^\mu=(N,\beta)$ are non-dynamical,
since their momenta vanish. 
A direct calculation (see the Appendix \ref{ApA}) gives 
the Hamiltonian density,
\be                      \label{Hs}
{\cal H}=N{\cal H}_0+\beta{\cal H}_r+{m^2}{\cal V}
(N,\beta,\Delta,R), 
\ee
where 
\bea                  \label{H0r}
{\cal H}_0&=&\frac{\Delta^3}{4R^2}\,\pii_\Delta^2+\frac{\Delta^2}{2R}\,\pii_\Delta\pii_R 
+\Delta (R^{\prime 2}+2RR^{\prime\prime})+2R\Delta^\prime R^\prime-\frac{1}{\Delta}\,,
\nonumber \\
{\cal H}_r&=&\Delta\pii_\Delta^\prime+2\Delta^\prime\pii_\Delta+R^\prime \pii_R\,.
\eea
These expressions have been much studied  (see for example  
\cite{Unruh:1976db,*Kuchar:1994zk}).
Setting $m=0$, General Relativity is recovered,
in which case varying the Hamiltonian with respect to 
$N,\beta$ gives two constraints: 
${\cal H}_0=0$ and  ${\cal H}_r=0$. 
These constraints are first class (see the Appendix \ref{ApE}),
hence they generate diffeomorphisms in the $t,r$ space,
which can be used to impose two gauge conditions on the 
phase space variables. As a result, there remain 
$4-2-2=0$ independent phase space variables, 
in agreement with the well-known fact that 
in vacuum General Relativity there is no dynamic in 
the s-sector (Birkhoff theorem). 

If ${m\neq 0}$ and the potential 
${\cal V}$ has the generic form \eqref{PF} (see Eq.\eqref{V} in the Appendix \ref{ApB}), 
then varying ${\cal H}$ with respect to 
$N,\beta$ does not give constraints but rather equations, 
\be                        \label{eqq}
{\cal H}_0+m^2\,
\frac{\partial {\cal V}(N,\beta,\Delta,R)}{\partial N}=0,~~~
{\cal H}_r+m^2\,
\frac{\partial {\cal V}(N,\beta,\Delta,R)}{\partial \nu}=0,~~~ 
\ee
which can be resolved for
$N=N(\Delta,R)$ and $\beta=\beta(\Delta,R)$. 
Since there are no constraints, 
all four phase space variables are independent
and describe two propagating DOFs. 
One of them can be associated with 
the scalar polarization of the massive graviton, 
while the other one should be attributed 
to the BD ghost. Inserting $N=N(\Delta,R)$ and $\beta=\beta(\Delta,R)$ into 
${\cal H}=N{\cal H}_0+\beta{\cal H}_r+{m^2}{\cal V}$ gives a function
that is unbounded from below. 

Let us now consider the dRGT theory, where (see the Appendix \ref{ApB}) 
\bea                    \label{XPEH}
{\cal V}=\sqrt{\h}N{\cal U}=\frac{NR^2}{\Delta}P_0
+\frac{R^2}{\Delta}P_1\sqrt{(N\Delta+1)^2-\beta^2}+R^2 P_2 \,,
\eea
with 
\be                                   \label{P}
P_m={b_m}+2b_{m+1}\frac{r}{R}
+{b_{m+2}}\frac{r^2}{R^2}
~~~~~~(m=0,1,2).
\ee
Equations \eqref{eqq} then read 
\bea                          \label{vars}
&&
{\cal H}_0+m^2\frac{R^2P_0}{\Delta}
+m^2R^2P_1\,\frac{N\Delta+1}{\sqrt{(N\Delta+1)^2-\beta^2}}=0\,,\nonumber \\
&&
{\cal H}_r
-m^2\frac{R^2P_1}{\Delta}\,
\frac{\beta}{\sqrt{(N\Delta+1)^2-\beta^2}}=0.
\eea
The second of these conditions can be resolved
with respect to $\beta$,
\be                    \label{nu}     
\beta=(N\Delta+1)\frac{\Delta{\cal H}_r}{Y}\,,
\ee
with
\be                             \label{Y}
Y\equiv \sqrt{(\Delta {\cal H}_r)^2+({m^2}R^2P_1)^2}\,.
\ee
Inserting this into the first relation in 
\eqref{vars} does not give an equation for $N$
but a constraint,  
\be                                 \label{CON_C}
{\cal C}\equiv{\cal H}_0+Y+{m^2}\frac{R^2P_0}{\Delta}=0,
\ee
while $N$ remains undetermined. Inserting 
\eqref{nu} into ${\cal H}=N{\cal H}_0+\beta{\cal H}_r+{m^2}{\cal V}$
gives
\bea                 \label{Hmm}
{\cal H}&=&
{\cal E}+N{\cal C} \,,
\eea
with 
\bea                            \label{Enn}
{\cal E}&=& \frac{Y}{\Delta}+{m^2}R^2 P_2\,,
\eea
so that 
varying ${\cal H}$ with respect to 
$N$ reproduces the constraint equation 
${\cal C}=0$ once again. 
Therefore, when restricted  to the constraint surface, 
${\cal E}$ in \eqref{Enn} gives the energy density. 
In what follows it will be convenient to use also
an equivalent representation 
for ${\cal H}$, 
\be                                 \label{Hmm0}
{\cal H}={\cal E}_{\rm 0}+\left(N+\frac{1}{\Delta}\right)\,{\cal C},
\ee
where 
\bea                            \label{En}
{\cal E}_{0}&=&{\cal E}-\frac{{\cal C}}{\Delta}=
-\frac{{\cal H}_0}{\Delta}
+{m^2}R^2 \left(P_2-\frac{P_0}{\Delta^2}\right),
\eea
which 
coincides  with 
 ${\cal E}$ on the constraint surface.

Since the constraint should be preserved in time, its Poisson 
bracket with the Hamiltonian 
\be 
{H}=\int_0^\infty {\cal H}\,dr
\ee
should vanish. It turns out 
that the constraint commutes with itself 
(see the Appendix \ref{ApE}),
\be 
\{{\cal C}(r_1),{\cal C}(r_2)\}_{\rm PB}=0,
\ee
therefore 
\be                           \label{S_PB}
{\cal S}\equiv \{{\cal C},{H}\}_{\rm PB}=0
\ee
is a new constraint since 
the term proportional to $N$ 
drops out of the bracket. A straightforward 
(but lengthy) computation of the bracket in \eqref{S_PB} 
uses the rules described  in the Appendix \ref{ApE} and gives 
\bea                                \label{CON_S}
{\cal S}&=&
\frac{m^4R^2P_1^2}{2Y}\,
(\Delta\pii_\Delta+R\pii_R)
-
\frac{\Delta^2\pii_\Delta}{2R}
\left\{
\frac{m^4}{2\Delta Y}\,
\partial_R (R^4P_1^2)+
m^2\partial_R (R^2P_2)
\right\} \nonumber \\
&-&
\frac{m^2{\cal H}_r}{Y}
\left\{
\Delta^2\left(R^2P_2\right)^\prime 
+R^2\partial_r(P_0-\Delta^2 P_2)
\right\}
-Y\left(\frac{\Delta{\cal H}_r}{Y}\right)^\prime\,.
\eea
Here the prime denotes the total 
derivative with respect to $r$, while 
$\partial_R$ and $\partial_r$ are the partial 
derivatives with respect to $R$ and $r$. 
It is worth noting that the two constraints have been known up to now only
implicitly \cite{Hassan:2011hr,*Hassan:2011ea,*Kluson:2012wf,*Comelli:2012vz,*Comelli:2013txa}, 
whereas Eqs.\eqref{CON_C} and \eqref{CON_S} provide explicit 
expressions for and values of the parameters $b_k$. 
Requiring further that 
$\{{\cal S},{H}\}_{\rm PB}=0$ gives 
an equation for $N$ because 
the two 
constraints do not commute with each other and 
the term proportional to $N$ does not drop out.
This equation is rather lengthy and will not be 
explicitly shown, unless for the special case described below 
in Section \ref{Sec6}. 

The two constraints remove one of the two DOFs. If 
the remaining DOF corresponds to  
the scalar graviton, then  
the energy should be positive.  
The  energy is
\be                      \label{Etot}
E=\int_0^\infty {\cal E}(\Delta,R,\pii_\Delta,\pii_R)\,dr=
\int_0^\infty {\cal E}_{0}(\Delta,R,\pii_\Delta,\pii_R)\,
dr\,,
\ee
where 
$\Delta,R,\pii_\Delta,\pii_R$ 
should fulfill two constraint equations
\be                  \label{CCC}
{\cal C}(\Delta,R,\pii_\Delta,\pii_R)=0,~~~~~
{\cal S}(\Delta,R,\pii_\Delta,\pii_R)=0.
\ee
These are non-linear ordinary differential equations, 
whose solutions $\Delta(r)$, $R(r)$, 
$\pii_\Delta(r)$, $\pii_R(r)$ 
can be used to describe initial data for the dynamical 
evolution problem. These equations
are rather complicated, but they simplify in some 
 cases.

\section{Weak field limit \label{Sec4}}
\setcounter{equation}{0}
 
In flat space, where 
$\Delta=1$, $R=r$, $\pii_\Delta=\pii_R=0$ 
and 
$b_k=b_k(c_3,c_4)$ (see Eq.\eqref{bbb}),
one has 
\be
{\cal C}={\cal S}={\cal E}={\cal E}_{0}=\beta=0,\,~~~N=1.
\ee
Let us consider the limit where 
the deviations
from flat space, 
\be 
\nu=N-1,~~~\beta,~~~ 
\delta=\Delta-1,~~~ \rho=R-r,~~~
\pii_\Delta,~~\pii_R\,,
\ee
are small. 
As shown in the Appendix \ref{ApC}, 
expanding  
the Hamiltonian density 
${\cal H}$ in Eq.\eqref{Hs} 
gives 
\be                                 \label{HFP}
{\cal H}={\cal E}_{\rm FP}
+\nuu\,{\cal C}_{\rm FP}+\mbox{cubic and higher order terms},
\ee 
where 
\bea                
{\cal E}_{\rm FP}= 
\frac{\pii_\Delta^2}{4r^2}+\frac{\pii_\Delta \pii_R}{2r}
+\frac{(\pii_\Delta^\prime +\pii_R )^2}{m^2 r^2}
+
2\rho\,\delta^\prime-\rho^{\prime 2}-\delta^2
+m^2(
2r\delta\rho-\rho^2),
                     \label{Hfp1B}
\eea 
and  
\be                          \label{C_FP}
{\cal C}_{\rm FP}=
(2r(\delta+\rho^\prime))^\prime+
m^2(r^2\delta -2r\rho).
\ee 
Truncating the higher order terms gives the FP Hamiltonian density, 
\be                                 \label{HFPa}
{\cal H}_{\rm FP}={\cal E}_{\rm FP}
+\nuu\,{\cal C}_{\rm FP},
\ee 
so that ${\cal E}_{\rm FP}$ is 
the FP energy density, while ${\cal C}_{\rm FP}=0$ is the constraint. 
Its preservation gives rise to 
the secondary constraint, 
${\cal S}_{\rm FP}\equiv
\{{\cal C}_{\rm FP},\int {\cal H}_{\rm FP}\,dr\}_{\rm PB}=0$, where   
\be                          \label{S_FP}
{\cal S}_{\rm FP}=
\frac{m^2}{2}\,(r\pii_R-\pii_\Delta)-
(\pii_\Delta^{\prime}+\pii_R)^\prime\,.
\ee
Therefore, the  energy 
in the weak field
limit is  
\be                                  \label{EE_FP}
E_{\rm FP}=\int_0^\infty{\cal E}_{\rm FP}\, dr\,,
\ee
where the arguments of ${\cal E}_{\rm FP}$ 
should fulfill the two constraints. As shown in the 
Appendix \ref{ApC}, the same results can be obtained by
expanding the energy ${\cal E}_0$  and constraints
${\cal C,S}$ given by Eqs.\eqref{CON_C},\eqref{En},\eqref{CON_S} 
from the previous Section. 
Therefore, the energy density ${\cal E}_0$ 
\eqref{En} agrees in the weak field limit with the FP 
energy density \eqref{Hfp1B}.

One can check that the FP energy \eqref{EE_FP} is positive. 
The first step is to resolve the constraints.
Introducing a new function $\xi=\delta+\rho^\prime$, the ${\cal C}_{\rm FP}=0$ constraint  
reduces to 
\be
\left(m^2r^2\rho-2r\xi\right)^\prime=m^2r^2\xi\,,
\ee
which is solved by 
\be
m^2r^2\rho-2r\xi=Q,~~~~m^2r^2\xi=Q^\prime
\ee
for an arbitrary $Q$. A similar trick works for the ${\cal S}_{\rm FP}=0$ constraint.
As a result, the constraints are solved by  
\bea                          \label{FP_c}
\delta&=&-\rho^\prime+\frac{Q^\prime}{r^2},~~~~~~
\rho=\frac{Q}{r^2}+\frac{2Q^\prime}{m^2r^3}, \nonumber \\
p_R&=&-p_\Delta^\prime+\frac{F^\prime}{r},~~~~~~
p_\Delta=\frac{F}{r}-\frac{2F^\prime}{m^2r^2}\,,
\eea 
where $Q,F$ are arbitrary functions. Inserting this into \eqref{Hfp1B} gives 
\be                      \label{T-FP}
{\cal E}_{\rm FP}=
\frac{3}{r^4}\left(
Q^{\prime 2}+m^2 Q^2+\frac{F^2}{4}
\right)+X^\prime,
\ee
with 
\be
X=\frac{2\rho Q^\prime}{r^2}-2\rho\rho^\prime
-m^2 r\rho^2-\frac{p_\Delta^2}{4r}
+m^2\,\frac{Q^2}{r^3}+\frac{F^2}{4r^3}\,.
\ee
In the weak field limit the energy must be finite and its density 
must be bounded. This requires  that  for $r\to 0$ the functions $Q$ and $F$
should approach zero faster than $r^{9/2}$ and $r^{7/2}$, respectively, while for 
$r\to\infty$ they should not grow faster than $r^{3/2}$. These conditions imply that 
the function $X$ vanishes for $r\to 0,\infty$, 
therefore the second term in \eqref{T-FP} does not contribute to the 
energy integral, while the first term in \eqref{T-FP} is non-negative.
Therefore, $E_{\rm FP}=\int_0^\infty{\cal E}_{\rm FP}\geq 0$,
in agreement with the general analysis in Appendix \ref{ApFP}.

\section{Arbitrary fields -- kinetic energy sector \label{Sec5}}
\setcounter{equation}{0}

Let us choose $b_k=b_k(c_3,c_4)$ according to \eqref{bbb}
and set $\Delta=1$ and  $R=r$,   
so that the 3-metric is flat. At the same time, 
the momenta $\pii_\Delta$, $\pii_R$ are allowed to assume any values. 
The polynomials $P_m$ 
defined by \eqref{P} then become $P_1=-P_0=-P_2=1$, and 
the energy \eqref{En}    
\be 
{\cal E}_{0}=
-\frac{(2r\pii_R+\pii_\Delta)\,\pii_\Delta}{4r^2}\,.
\ee
The energy is carried only by the momenta, so it is purely kinetic, but 
it is not obvious that it is positive. 
The constraint \eqref{CON_C} becomes 
\be 
{\cal C}=
\frac{(2r\pii_R+\pii_\Delta)\,\pii_\Delta}{4r^2}
+
\sqrt{(\pii_\Delta^\prime+\pii_R)^2+m^4r^4}
-m^2r^2=0,\,
\ee
while the secondary constraint \eqref{CON_S} reduces to a rather lengthy expression,
\bea
{\cal S}&=&\left\{\frac{m^2}{2}\,(r\pii_R-\pii_\Delta)
-(\pii_\Delta^\prime+\pii_R)^\prime\right.\nonumber  \\
&+&\frac{(2r\pii_R+\pii_\Delta)\,
\pii_\Delta\pii_\Delta^{\prime\prime}}{4\,m^2\,r^4}
-\frac{(r\pii_R+\pii_\Delta)\,
\pii_\Delta^{\prime 2}}{m^2r^4}
+\frac{(\pii_\Delta^2-3r^2\pii_R^2-r^2\pii_\Delta\pi_R^\prime
-2r\pii_\Delta\pii_R)\,
\pii_\Delta^{\prime}}
{2m^2r^5}  \nonumber \\
&+&\frac{\pii_\Delta^2\pii_R^\prime}{4\,m^2\,r^4}
+\frac{\pii_R(\pii_\Delta^2-r^2\pii_R^2)}{2m^2r^5}
+\frac{\pii_\Delta(\pii_\Delta+2r\pii_R)((c_3-2)\pii_\Delta-r\pii_R)}{4r^4}
\nonumber \\
&+&\left.\frac{(r\pii_R+\pii_\Delta)
(2r\pii_R+\pii_\Delta)\pii_\Delta^2}
{32m^2r^8}\right\}
\left\{
1-\frac{(2r\pii_R+\pii_\Delta)\pii_\Delta}{4m^2r^4}
\right\}^{-1}=0\,.
\eea
If $\pii_\Delta,\pii_R$ are small,
then 
\bea 
{\cal C}&=&\frac{(2r\pii_R+\pii_\Delta)\,\pii_\Delta}{4r^2}
+\frac{(\pii_\Delta^\prime+\pii_R)^2}{2m^2r^2}+\ldots,
\nonumber \\
{\cal S}&=&{\cal S}_{\rm FP}+\ldots,~~~~\nonumber \\
{\cal E}_{0}+2{\cal C}&=&{\cal E}_{\rm FP}+\ldots,
\eea
so that the FP limit is recovered. 
The first constraint can be represented in the form 
\be                         \label{CC1}
\left(\frac{(2r\pii_R+\pii_\Delta)\,\pii_\Delta}{4r^2}
-m^2r^2\right)^2
={(\pii_\Delta^\prime+\pii_R)^2+m^4r^4}\,.
\ee
Differentiating this yields  an expression 
for  $\pii_\Delta^{\prime\prime}$, which
can be used to remove the second derivative from 
${\cal S}$. In addition, Eq.\eqref{CC1}
can be used to remove also  
$\pii_\Delta^{\prime 3}$ and $\pii_\Delta^{\prime 2}$. 
As a result,
the second constraint simplifies and reduces to 
\be                                \label{SS1}
\pii_\Delta^2(\pii_\Delta+2r\pii_R)
[2(c_3-1)\,r\pii_\Delta^\prime 
+2(c_3-2)\,r \pii_R-\pii_\Delta
]+4\,m^2r^6(\pii_\Delta\pii_R^\prime 
+\pii_R^2+2\pii_\Delta^\prime\pii_R)=0.
\ee
Further simplifications can be  achieved via
passing to the dimensionless 
radial coordinate $x=mr$ and 
expressing the two momenta in terms of two new function $z,f$ as 
\be 
\pii_\Delta=\frac{\sqrt{xz}}{m}\,,~~~~
\pii_R=-\frac{(xz+4x^4 f)}{2x\sqrt{xz}}\,.
\ee
With these definitions Eqs.\eqref{CC1},\eqref{SS1} reduce to 
\bea                          \label{zf}
\frac{dz}{dx} &=&4\,x^2f+
2x\sqrt{xz}\,F\,, \\
\frac{df}{dx} &=&\frac{
4\,(1-c_3)\,zf
-4x^3f-3z
}{4x\sqrt{xz}}\,F
-\frac2x\,F^2\,,\nonumber 
\eea
with $F=\pm\sqrt{f(f+2)}$ (nothing depends on $c_4$). 
The energy density is 
\be
{\cal E}_0=x^2f\,.
\ee
Since $F^2=f(f+2)\geq 0$, one has either   
$f\geq 0$ or $f\leq -2$, which
determines two different solution branches whose energy is either
non-negative or strictly negative.   
There can be no interpolation 
between these branches since this would require crossing the region of 
forbidden values of $f$.

A simple solution from the first branch is $f=0$, $z=z_0$, whose energy is 
zero. It reduces to the flat space configuration for $z_0=0$. 
If the solutions of Eqs.\eqref{zf} are to describe initial values 
for perturbations around flat space,
then 
they should correspond to smooth deformations of the latter,
and this selects the $f\geq 0$ branch. Therefore, 
the energy for perturbations around flat space 
is positive.  

 A simple solution from the second branch is 
$f=-2$ and $z=\frac{8}{3}\,(x_{\rm max}^3-x^3)$, 
where $x_{\rm max}$ is an integration constant. 
Since $z$ should be positive, the solution exists only
for $x\leq x_{\rm max}$, with 
the  energy 
$E=\int_0^{x_{\rm max}}{\cal E}_0\,dr=-\frac{2}{3m}\,x_{\rm max}^3$. 
As $x_{\rm max}$ can be arbitrarily large, the energy is unbounded 
from below.

One can construct more general negative energy solutions of Eqs.\eqref{zf}
numerically. They typically exist 
only within a finite interval of $x$,
because either $f\to -\infty$ of $z\to 0$ at the
ends of the interval.
Such solutions cannot describe regular initial data and they  
belong to the disjoint from flat space branch. 
Therefore, they cannot affect 
the stability of flat space. 

Summarizing, the energy can be negative and even unbounded from below,
but only in a disconnected from flat space sector, while  
the energy for smooth excitations over flat space is positive.


\section{Arbitrary fields -- potential energy sector \label{Sec6}}

Let us now set the momenta to zero, 
$\pii_\Delta=\pii_R=0$, allowing at the same time
the metric coefficients  $\Delta$ and $R$ to vary. 
Since the momenta are trivial, the kinetic energy vanishes, 
but there remains the potential energy of metric deformations. 
The second constraint is trivially satisfied for zero momenta, 
$
{\cal S}=0.
$
Eq.\eqref{Y} yields  $Y=m^2R^2P_1$ and 
the first constraint becomes 
\be
{\cal C}=
\Delta (R^{\prime 2}+2RR^{\prime\prime})+2R\Delta^\prime R^\prime-\frac{1}{\Delta}
+m^2R^2\left(P_1+\frac{P_0}{\Delta}\right)=0,
\ee
while the energy density \eqref{Enn} is 
\be 
{\cal E}=m^2R^2\left(P_2+\frac{P_1}{\Delta} \right).
\ee
It is convenient to set
\be 
\Delta=\frac{g(r)}{h(r)}\,,~~~~~R=rh(r).
\ee
Choosing 
$b_k=b_k(c_3,c_4)$ according to \eqref{bbb}, the constraint reduces to  
\bea 
&-&h^{\prime\prime}-\frac2r\,h^\prime+
\frac{h^{\prime 2}}{2h}-\frac{(rh)^\prime g^\prime}{rg}
+\frac{h(1-g^2)}{2r^2g^2} \\
&+&m^2\,\frac{(6-4c_3-c_4)h^3
+(2c_4+6c_3-6)h^2+(1-2c_3-c_4)h}{2g^2} \nonumber \\
&+&m^2\,\frac{(c_4+3c_3-3)\,h^2
+(2-4c_3-2c_4)\,h+c_3+c_4}{2g}
=0,\nonumber 
\eea
while 
\be 
{\cal E}=-\left.\left.\frac{m^2r^2}{g}\,
\right((c_4+3c_3-3)\,h^3
+(1-2c_3-c_4)(g+2)\,h^2
+(c_3+c_4)(1+2g)\,h-c_4g
\right).
\ee
The simplest solutions of the constraint 
are obtained by setting $g(r)=1$
and $h(r)=h_0$, which gives for $h_0$ 
an algebraic equation 
with three roots,
\be 
h_0=\left\{1,\frac{3-5\,c_3-2\,c_4
\pm \sqrt{(3\,c_3+1)^2+12\,c_4+8}}
{6-4\,c_3-c_4}\right\}.
\ee
For the first root, $h=1$, one has 
${\cal E}=0$, while for the two others 
one has ${\cal E}={\rm const.}\times m^2r^2$,
where the constant can be  positive
or negative. 
For example, for $c_3=0.1$, $c_4=0.3$
the roots and the corresponding energies, respectively, are 
\be 
h_0=\left\{1,-0.14,
0.50\right\},~~~~
{\cal E}=\left\{0,+0.43\,m^2r^2 ,
-0.38\, m^2r^2\right\}.~~~~
\ee 
Therefore, the energy density can be 
positive or negative.  
Solutions with $h_0\neq 1$ are 
globally regular but 
non-asymptotically flat;  their total 
energy is infinite and can be positive or negative. 
As a result, one can see again that the energy is unbounded from below.

Let us set for simplicity $c_3=c_4=0$
and pass to the dimensionless variable 
$x=mr$. The prime from now on will denote the derivative 
with respect to $x$. 
The constraint reduces to 
\bea                           \label{eqC0}
h^{\prime\prime}
+\frac2x\,h^\prime-
\frac{h^{\prime 2}}{2h}+\frac{(xh)^\prime g^\prime}{xg}
-\frac{h(1-g^2)}{2x^2g^2} 
-\frac{h(2-3h)}{2g} 
-\frac{h(1-6h+6h^2)}{2g^2}
=0,
\eea
while the energy density 
\bea                  \label{eqE}
{\cal E}&=&\frac{x^2h^2(3h-g-2)}{g}\,.
\eea
Expressing $g(x)$ in terms of a new function $q(x)$ as
\be
g=\frac{qh}{(xh)^\prime}\,,
\ee
the constraint  becomes 
\be                           \label{eqC00}
\left\{xh(1-q^2)+x^3 h(h-1)(2h-1)\right\}^\prime=
x^2 h(q-1)(3h-2), 
\ee
which is equivalent to 
\bea                           \label{eqC}
Q&=&xh(1-q^2)+x^3 h(h-1)(2h-1), \nonumber \\
Q^\prime&=&x^2 h(q-1)(3h-2),
\eea
with an arbitrary function $Q(x)$. For any chosen
$Q$ equations \eqref{eqC} can be algebraically resolved   
with respect to $h$  and $q$, which gives a solution of the constraint. 

Even though the second constraint is trivially satisfied, 
the condition of its preservation, 
 $\{{\cal S},H\}_{\rm PB}=0$, is non-trivial and reduces to 
$
{\cal A}N-{\cal B}=0\,,
$ 
where 
\bea
{\cal A}&=&q(\alpha_1+\alpha_2)+2(q-1)^2(27h^2-18h+4)(xh)^\prime 
-6x^2h(3h-1)(3h-2)(4h-3)(xh)^\prime, \nonumber  \\
{\cal B}&=&(\alpha_2-\alpha_1)(xh)^\prime+8h^2q(q-1)^2
+6h^2x^2q(3h-2)^2\,,
\eea
with 
\be 
\alpha_1=3x^2h(3h-2)(13h^2-12h+2),~~~~~
\alpha_2=2h(q-1)^2(9h-2).
\ee
Therefore, the lapse function is 
$
N={\cal B}/{\cal A},
$
while 
the shift function obtained from Eq.\eqref{nu} is $\beta=0$.  
The 3-metric will be regular 
and asymptotically flat 
if $h$ and $q$ are smooth and fulfill the  boundary conditions 
\be                  \label{bc}
h_0\leftarrow h \to 1,~~
1\leftarrow q \to 1~~~~~\mbox{for}~~
0\leftarrow x \to \infty,
\ee
with $h_0>0$. The simplest solutions of the constraint 
are obtained by setting in \eqref{eqC} $Q=0$,
which implies that $q=1$ but yields  three different solutions for $h$,
\be                              \label{h}
h(x)=\left\{1,\frac12,0\right\}~~~~\Rightarrow
~~~~~{\cal E}(x)=\left\{0,-\frac38\,x^2,0\right\}.
\ee  
Interestingly, these solutions of the constraint fulfill also 
the complete system of the Hamilton equations since one has for them 
\be
\dot{\Delta}=\{\Delta,H\}_{\rm PB}=0,~~ 
\dot{R}=\{R,H\}_{\rm PB}=0,~~ 
\dot{p}_\Delta=\{p_\Delta,H\}_{\rm PB}=0,~~ 
\dot{p}_R=\{p_R,H\}_{\rm PB}=0.
\ee  
If $h=0$ then the metric is degenerate, which case is not interesting,
while the two other solutions in \eqref{h}  give rise 
to two different branches of regular  solutions of the constraint.

\subsection{Positive energy branch}
\begin{figure}[h]
\hbox to \linewidth{ \hss
	
	\resizebox{7cm}{6cm}
	{\includegraphics{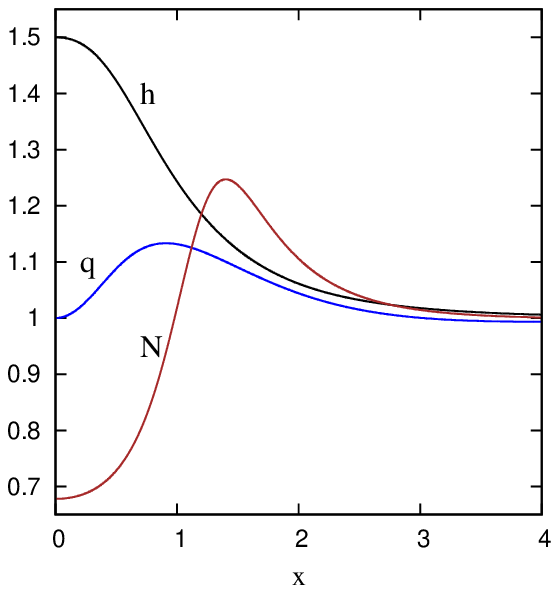}}
\hspace{1mm}
	\resizebox{7cm}{6cm}{\includegraphics{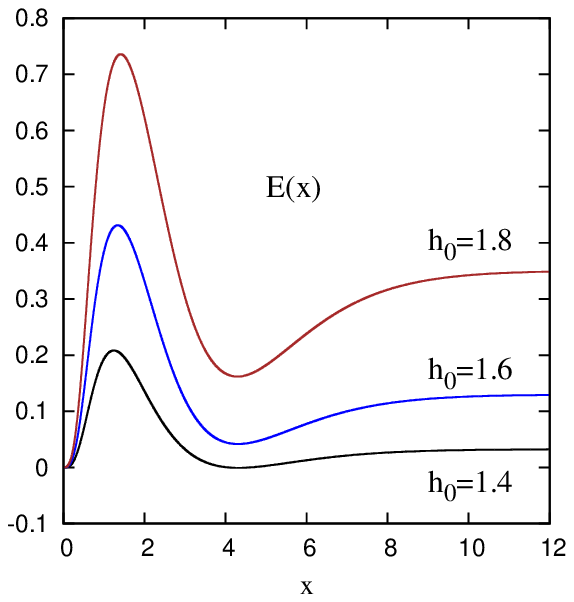}}
	
\hspace{1mm}
\hss}
\caption{{\protect\small 
Profiles of $h,q,N$,  and $E(x)$  for 
the positive energy solutions with $Q=kr^5/(1+x^2 e^x)$. 
 }}%
\label{Fig1}
\end{figure}

For the $h=1$ solution in \eqref{h} one has  
$N=1$ and the 4-metric is flat, $ds_g^2=\,ds_f^2$. 
The energy is zero.   Let us consider 
deformations of this solution 
by changing the value of $h$ at the origin. 
Eq.\eqref{eqC00} then yields  
\be
h=h_0+{\cal O}(x^2),~~~~ 
q=1+{\cal O}(x^2)
\ee for small $x$, in which case 
Eqs.\eqref{eqC} require that 
\be                                   \label{k}
Q=kr^5+{\cal O}(x^7)~~~\mbox{with}~~~
k=\frac{1}{10}\,h_0(2h_0-1)(h_0-1)(3h_0-2).
\ee 
This suggests that one can choose the function $Q$, for example, as
\be
Q=\frac{kr^5}{1+Ax^2 e^x},
\ee
where $A$ is a parameter. Inserting this to 
\eqref{eqC} and resolving with respect to 
$h$ and $q$ gives the globally regular and asymptotically flat 
 solutions shown in Fig.\ref{Fig1}. 

These solutions describe 
smooth metric deformations of the flat space. 
They correspond only to the initial time moment, 
since later the metric will dynamically evolve,
and to determine its temporal evolution will require 
solving the full system of Hamilton equations.  
However, the total energy computed at the initial time moment
will be the same for all times, and, as can be seen in Fig.\ref{Fig1},
is positive.  
Specifically, 
the energy  contained in the sphere or radius $x$ (expressed in $1/m$ units), 
\be
E(x)=\int_0^x {\cal E}\, dx, 
\ee
can be negative for small $x$ (if $h_0<1$), but the total energy 
$E(\infty)$ turns out to be always positive and grows 
when $|h_0-1|$ increases. 
As a result, the energy is positive for smooth, asymptotically flat 
fields, so that the 
positivity of their energy in the weak field limit 
holds in the fully non-linear theory as well. 

\subsection{Tachyon branch}
For the $h=\frac12$ solution in \eqref{h} 
one has $N=\frac12$ and  the  two 
metrics are proportional, $ds_g^2=\frac14\,ds_f^2$.
Even though they are both flat, 
this solution is quite different from flat space
since one now has  $E(x)=-x^3/8$, which corresponds to the constant and 
negative energy density. The total energy is negative and infinite.

For small fluctuations around this background one has 
\be
\Delta=2+\delta,~~~~R=\frac{r}{2}+\rho,
\ee
in addition the momenta $p_\Delta,p_R$ are non-zero but small. 
Linearizing the  constraints \eqref{CON_C},\eqref{CON_S} with respect to small
$\delta,\rho,p_\Delta,p_R$ then gives 
the FP constraints \eqref{C_FP},\eqref{S_FP}, up to the replacement
\be 
m^2\to -\frac{m^2}{2}.
\ee
Therefore, the FP mass becomes imaginary for fluctuations around 
this background, 
hence gravitons become tachyons. 

One can also construct more general solutions 
by setting in \eqref{k} $h_0\approx \frac12$, in which case 
$h(x)\to\frac12$ as $x\to\infty$. 	 
The total energy is always 
negative and infinite, which can be viewed as an indication  
of the presence of the ghost. 
However, if the tachyon branch is completely disjoint 
from the positive energy branch, then the ghost will be harmless, 
since it will not be able to affect the positive energy states.

\subsection{Tachyon bubbles.}
\begin{figure}[h]
\hbox to \linewidth{ \hss

	\resizebox{7cm}{6cm}
	{\includegraphics{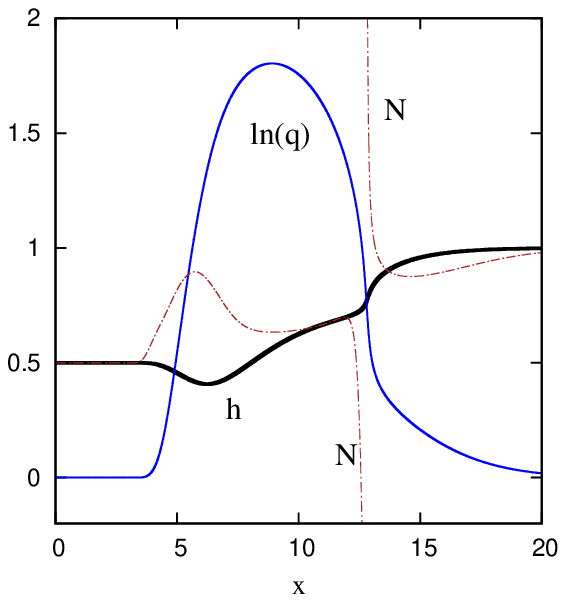}}
\hspace{1mm}
	\resizebox{7cm}{6cm}{\includegraphics{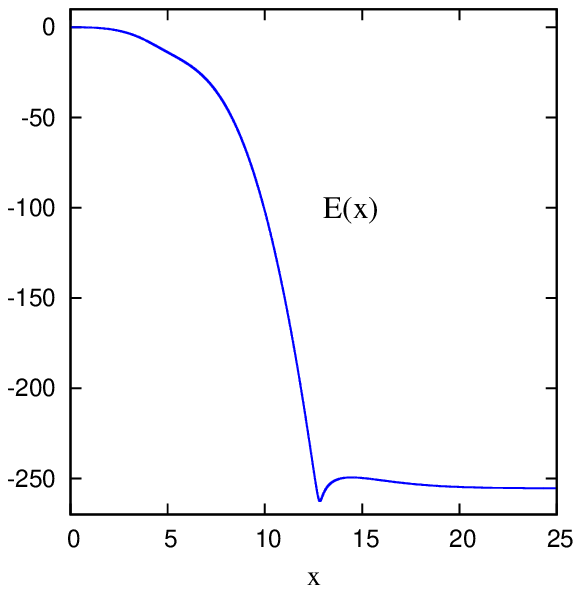}}
	
\hspace{1mm}

\hss}

\caption{{\protect\small Profiles of $h$ (thick line), 
$\ln(q)$, $N$, and $E(x)$ for the tachyon bubble solution
with  $Q=-\Theta(x-3)(x-3)^8 e^{-x}$.
 }}%
\label{Fig2}
\end{figure}

It is not immediately obvious that the tachyon branch is disjoint 
from the positive energy branch since 
there are solutions which interpolate 
between the two. For these solutions one has $h=1/2$ at the origin
but $h\to 1$ at infinity; they can be obtained by 
choosing in \eqref{eqC}  
\be 
Q=A\,\Theta(x-x_0)(x-x_0)^p e^{-x},
\ee 
where $\Theta(x)$ is the step function and $p$ is positive and large enough. 
Such a choice of $Q$ 
 enforces for $h$ a kink-type behaviour, so that 
$h=\frac12$ for $x<x_0$ but $h$ starts to grow for $x>x_0$  and $h\to 1$
as $x\to\infty$ (see Fig.\ref{Fig2}). Solutions thus 
start from the tachyon phase at the origin but 
approach flat space at infinity, so that they describe 
bubbles of the tachyon phase of size $\propto x_0$. 
If $x_0$ is large, then the energy 
$E\propto -x_0^3$ (see Fig.\ref{Fig2}).

The bubble 3-metric is  regular and asymptotically flat, while the energy is negative. 
This is embarrassing, 
since this suggests  that the flat space
could decay into bubbles. 
However, a more close inspection reveals that 
the lapse function $N$ for the bubbles 
is necessarily singular. 
Indeed, one has $N={\cal B}/{\cal A}$, but ${\cal A},{\cal B}$
are both negative for $h=\frac12$ and become
positive for $h=1$, hence each of them vanishes 
at least once as $h$ interpolates between $\frac12$ and $1$. 
Next, $h$ must 
cross the value $h=2/3$ at some point where $h^\prime>0$.  
Assuming regular Taylor expansions for $h$ and $q$ and 
constructing the power-series solution of the constraint \eqref{eqC00} at this point,
it turns out that ${\cal A}$ and ${\cal B}$ are both negative there. 
Therefore, they must change sign in the region where $h>2/3$.   
Next, one should check if they can vanish simultaneously. For this, 
one constructs a power-series solution of the constraint at a point
where  $x=x_0>0$,  $h=h_0>2/3$, and $q=q_0>0$, and one imposes on 
this solution two additional conditions, ${\cal A}={\cal B}=0$. 
This yields an algebraic equation 
for $x_0,h_0,q_0$, and it turns out that this equation has no solutions. 
As a result, ${\cal A},{\cal B}$ cannot vanish simultaneously. Therefore, 
$N$ must have at least one zero and a pole, as shown in Fig.\ref{Fig2}.

Since $N$ enters the Hamilton equations 
$\dot{p}_k=\{p_k,H\}_{\rm PB}$, the time derivative of the momenta 
diverges where $N$ has pole(s). Therefore, the bubble solutions do not 
describe regular initial data. 
It follows that the negative energy branch 
is totally disjoint from the positive energy branch
so that it cannot affect the stability of  flat space. 

The above conclusions apply to the theory with $c_3=c_4=0$, but 
the tachyon  bubbles can be constructed  also for $c_3\neq 0$ and $c_4\neq 0$.
The analysis then becomes more complicated and 
the above analytical arguments showing that the lapse function $N$ 
must be singular do not directly apply.
Nevertheless, the problem can be tackled numerically, and in all studied 
cases the lapse 
$N$ is found to be singular and even worse -- 
when one varies $c_3$ and $c_4$ or the function $Q(x)$, 
the lapse $N$ generically 
starts to exhibit many poles instead of just one pole.

\section{Conclusions --   Stability of the theory \label{Sec7}} 

To recapitulate the above discussion, 
the energy in the $s$-sector of the dRGT theory is found to be positive  for 
globally regular and asymptotically flat fields. Besides, 
there are also solutions of the constraints for which the energy 
can be negative and even unbounded from below and for which 
the gravitons behave as tachyons. 
The negative energies and tachyons can clearly be interpreted 
as a very bad sign, supporting the viewpoint that the whole theory 
is sick 
\cite{Deser:2012qx,*Deser:2013eua,*Deser:2013qza,*Deser:2014hga}. 
However, it is interesting that a different interpretation 
is also possible, and this we shall now try to advocate.  

The main point is that the above {\it global} analysis of the 
constraints shows that their negative energy solutions are always 
either not globally regular or not asymptotically flat. Therefore,  
they cannot describe initial data for a decay of the flat space. 
This indicates that the existence of the negative energies 
in the theory could actually be harmless since it does not affect 
the stability of the flat space and of its globally regular deformations.

One can give the following interpretation.  
 Globally regular and asymptotically flat fields constitute the 
``physical sector'' of the theory where the energy is positive and the 
ghost is absent/bound. This sector is healthy. 
As for the negative energy states, 
they belong to different sectors 
separated from the physical sector by a potential barrier. 

One may wonder how high is the potential barrier between 
the sectors. To estimate, 
one can compute 
the energy for  an interpolating sequence of fields. For example, 
fields which fulfill the constraints and  
satisfy the boundary conditions  \eqref{bc} will interpolate 
between the normal and tachyon branches when the 
parameter $h_0$ in \eqref{bc} varies from $1$
to $1/2$. A numerical evaluation shows that when $h_0$ 
decreases from unit value, 
the energy rapidly grows
(since the function $g$ in the denominator in \eqref{eqE} develops a minimum),
then it passes through a  pole and finally   
approaches a finite negative value 
when $h_0$ tends to $1/2$. This indicates that the potential barrier between the two 
sectors is infinitely high.

These arguments support the viewpoint that the physical sector could be protected 
from the influence of the negative energies. Interestingly, they can be used to argue 
that the physical sector could be protected also from the 
waves propagating faster than light, whose existence in the dRGT theory was 
discovered by using 
the {\it local} analysis of the differential equations \cite{Deser:2012qx,*Deser:2013eua,*Deser:2013qza,*Deser:2014hga}.
Indeed, it is natural to expect the superluminal waves 
to coexist with the tachyons, but, as suggested by the above arguments based on the 
{\it global} analysis, the tachyons should decouple to disjoint sectors, as their energy is negative. 
In other words, it is possible that the superluminal waves cannot develop
starting from globally regular and asymptotically flat initial data, in which case they 
would not appear in the physical sector.  
Although not a proof, this indicates  that the physical sector 
could be protected from  
superluminalities and perhaps also  from other seemingly non-physical features \cite{Deser:2013rxa},
\footnote{
It is argued in \cite{Deser:2012qx,*Deser:2013eua,*Deser:2013qza,*Deser:2014hga} 
that the dRGT theory admits not only superluminal waves but also closed causal curves,
a least in the case where the two metric have different signatures. However, since 
$\sqrt{g^{\mu\sigma}f_{\sigma\nu}}$ should be complex-valued in that case, 
it is not quite clear if one can consistently set the two metric signatures 
to be not the same. 
}.

It should be emphasised at the same time that the  above interpretation can at best be viewed 
only as a conjecture, as it is currently  based only on the results of the $s$-sector analysis.  
Of course, these results are suggestive. 
Indeed, as the ghost is a scalar and can propagate in the $s$-sector,
this sector would be the most natural place for the instability to show up. 
Therefore, its absence  in the $s$-sector indicates
that it could be absent in all sectors. However, to 
really prove this would require demonstrating that the energy is positive  
for arbitrary globally regular deformations of the flat space and that the 
negative energies totally decouple.  
Such a demonstration is lacking at present. 
Therefore, despite the positive evidence 
mentioned above, the issue of weather the dRGT theory can indeed be considered 
as a consistent theory remains actually open
\footnote{
A ghost-type instability in the dRGT theory with flat reference metric 
was detected for perturbations around
the de Sitter space \cite{DeFelice:2012mx,*Fasiello:2013woa,*Chamseddine:2013lid}. 
However, it is unclear if this invalidates the whole theory, as there are 
infinitely many inequivalent versions of de Sitter solution 
in the theory \cite{Volkov:2013roa}, only one of them being considered in 
\cite{DeFelice:2012mx,*Fasiello:2013woa,*Chamseddine:2013lid}. 
}

It is interesting that within the bigravity generalization of the dRGT 
theory,   where both metrics are dynamical \cite{Hassan:2011zd},
the tachyon vacuum in \eqref{h} is no longer a solution, as it does not fulfill 
the equations for the second metric \cite{Volkov:2011an,*Volkov:2013roa}. 
Since there are no tachyons, one does not expect 
the superluminalities to be present either, and indeed their existence 
within the bigravity theory has not been reported \cite{Deser:2013gpa}. 
It seems therefore that the bigravity theory could be better 
defined  than the massive gravity since it contains less negative energy solutions,
or maybe no such solutions at all. However, a detailed  analysis 
is needed in order to make definite  statements.

\acknowledgments
It is a pleasure to acknowledge discussions with 
Eugeny Babichev, Thibault Damour, Cedric Deffayet, 
Claudia de Rham, and 
Andrew Tolley. 
This work was partly supported by the Russian Government Program of Competitive Growth 
of the Kazan Federal University.


\appendix

\section{General Relativity Hamiltonian in the s-sector \label{ApA}}

\renewcommand{\theequation}{\Alph{section}.\arabic{equation}}

Let us consider
the spherically-symmetric  
spacetime metric \eqref{sphere},  
\bea 
ds_g^2&=&-N^2dt^2+\frac{1}{\Delta^2}\,(dr+\beta\, dt)^2
+R^2\,(d\vartheta^2+\sin^2\vartheta d\varphi^2),    \label{A1} 
\eea 
where $N,\beta,\Delta,R$ depend on 
of $t,r$\,. This corresponds to the ADM 
decomposition \eqref{ADM} with the 3-metric
\be
\h_{ik}\,dx^i dx^k=\frac{dr^2}{\Delta^2}+R^2\,
(d\vartheta^2+\sin^2\vartheta d\varphi^2),
\ee
and with the shift vector $N^k=(N^r,N^\vartheta,N^\varphi)=(\beta,0,0)$.
One has $N_i=\h_{ik}N^k=(\beta/\Delta^2,0,0)$
and $\sqrt{\h}=R^2/\Delta$, while the 
curvature scalar for the 3-metric is 
\be 
-\frac12\,\sqrt{\h}\,
R^{(3)}=
2\Delta RR^{\prime\prime}+2\Delta^\prime RR^\prime
+\Delta R^{\prime 2}-\frac{1}{\Delta}.
\ee
Calculating the second fundamental form ($\nabla^{(3)}$ is the covariant derivative   
with respect to $\h_{ik}$), 
\be                             \label{A4}
K_{ik}=\frac{1}{2N}(\dot{\h}_{ik}-\nabla^{(3)}_i N_k-\nabla^{(3)}_k N_i),
\ee
gives  for 
$K^i_{~k}=\h^{im}K_{mk}$ 
the only non-trivial components 
\be                  \label{AK}
K^r_{~r}=-\frac{1}{N\Delta}\left(\dot{\Delta}
+\Delta\beta^\prime
-\beta\Delta^\prime
\right)\,,~~~~~
K^\vartheta_{~\vartheta}=K^\varphi_{~\varphi}
=\frac{1}{NR}\left(\dot{R}-\beta R^\prime\right).
\ee
The Lagrangian is 
\bea
{\cal L}& =&\frac12\,\sqrt{\h}N\left(K^i_{~k}K^{k}_{~i}-(K^k_{~k})^2+R^{(3)}
\right)-m^2{\cal V} \\
&=&
\frac{\dot{R}-\beta R^\prime}{N^2\Delta^2}
\left(
2R\dot{\Delta}-\Delta\dot{R}
+\beta\Delta R^\prime +2\beta^\prime \Delta R
-2\beta \Delta^\prime  R
\right)+\frac12\,\sqrt{\h}
R^{(3)}-m^2{\cal V}\,. \nonumber 
\eea
Choosing $\Delta,R$ to be the dynamical variables, their  
momenta are 
\bea
\pii_\Delta&=&\frac{\partial {\cal L}}{\partial \dot{\Delta}}=
\frac{2R(\dot{R}-\beta R^\prime)}{N\Delta^2}\,,~~~
\nonumber \\
\pii_R&=&\frac{\partial {\cal L}}{\partial \dot{R}}=
\frac{2(R\dot{\Delta}-\Delta\dot{R}
+\beta^\prime\Delta R-\beta\Delta^\prime R
+\beta\Delta R^\prime) }{N\Delta^2}\,,~~~
\eea
which relations can be inverted, 
\be                        \label{dots}
\dot{\Delta}=\frac{N\Delta^2}{2R^2}\,
(\Delta\,\pii_\Delta+R\,\pii_R)
+\Delta^\prime\beta-\Delta\beta^\prime\,,~~~~~~~~
\dot{R}=\frac{N\Delta^2}{2R}\,\pii_\Delta
+\beta R^\prime\,.
\ee
The Hamiltonian density ${\cal H}=\dot{\Delta}\pii_\Delta+\dot{R}\pii_R-{\cal L}$ reduces to 
\be                        \label{HamA}
{\cal H}=
N{\cal H}_0+\beta{\cal H}_r+m^2{\cal V}\,
\ee 
with 
\bea                  \label{H0rA}
{\cal H}_0=\frac{\Delta^3}{4R^2}\,\pii_\Delta^2+\frac{\Delta^2}{2R}\,\pii_\Delta\pii_R 
-\frac12\,\sqrt{\h}\,R^{(3)},~~~~
{\cal H}_r=\Delta\pii_\Delta^\prime+2\Delta^\prime\pii_\Delta+R^\prime \pii_R\,.
\eea
This gives rise to
Eq.\eqref{H0r} in the main text. 
One can equally apply
Eqs.\eqref{Hm1},\eqref{Hm2} from the main text,
according to which 
\be                          \label{Hm1A}
{\cal H}_0=\left.\left.\frac{1}{\sqrt{\h}}\,
\right(2\pi^{i}_{~k}\pi^{k}_{~i}-(\pi^{k}_{~k})^2
\right)-\frac12\,\sqrt{\h}R^{(3)},~~~
{\cal H}_r=-2\nabla^{(3)}_i \pi^i_r\,,
\ee
with
\be                               \label{Hm2A}
\pi^{i}_{~k}
=\frac12\,\sqrt{\h}\,(K^{i}_{~k}-K^m_{~m}\,\delta^i_k).
\ee
Using \eqref{AK},\eqref{dots}, 
the only non-vanishing momenta are 
\be 
\pi^{r}_{~r}=-\frac12\,\Delta\,\pii_\Delta\,,~~~~
\pi^\vartheta_{~\vartheta}=
\pi^\varphi_{~\varphi}=\frac14\,R\,\pii_R\,.
\ee
Inserting this to \eqref{Hm1A} again 
reproduces Eq.\eqref{H0rA}. The only
subtlety is that $\pi^i_{~k}$
is a tensor density, whose covariant derivative 
is $\nabla^{(3)}_i \pi^i_k=\sqrt{\h}\,
\nabla^{(3)}_i( \pi^i_k/\sqrt{\h})$, where 
$\pi^i_k/\sqrt{\h}$ is a tensor whose 
covariant derivative 
is computed in the usual way.

\appendix
\setcounter{section}{1}
\section{Metric potential in the s-sector\label{ApB}}

\renewcommand{\theequation}{\Alph{section}.\arabic{equation}}

Let us calculate the potential of the dRGT
theory given by Eq.\eqref{2} in the main text. 
The first step is to consider 
the inverse of the spacetime metric  \eqref{A1}, 
\be                                  \label{gam}
g^{\mu\nu}=\left(
\begin{array}{cccc}
-{1}/{N^2} & {\beta}/{N^2} & 0 & 0 \\
{\beta}/{N^2} & \Delta^2-{\beta^2}/{N^2} & 0 & 0 \\
0 & 0 & {1}/{R^2} & 0 \\
0 & 0 & 0 & {1}/{(R^2\sin^2\vartheta)}
\end{array}
\right)\,,
\ee
while the f-metric is 
\be 
ds_f^2=-dt^2+dr^2+r^2\,(d\vartheta^2+\sin^2\vartheta d\varphi^2),
\ee
and therefore 
\be                                  \label{gamB}
{g^{\mu\sigma}f_{\sigma\nu}}=\left(
\begin{array}{cccc}
1/N^2 & \beta/N^2 & 0 & 0 \\
-\beta/N^2 & \Delta^2-\beta^2/N^2 & 0 & 0 \\
0 & 0 & r^2/R^2 & 0 \\
0 & 0 & 0 & r^2/R^2
\end{array}
\right)\,.
\ee
Let us apply this first to calculate 
the potential \eqref{PF} with all higher order terms truncated,   
\be                            \label{PF:B}
{\cal U}= \frac18\,
(H^{\mu}_{~\nu}H^{\nu}_{~\mu}-(H^\mu_{~\mu})^2),
\ee 
where $H^\nu_\mu=\delta^\mu_\nu-g^{\mu\sigma}f_{\sigma\nu}$.
With $u\equiv r/R$ 
one obtains for ${\cal V}=\sqrt{\h}N{\cal U}$
\be                    \label{V}
{\cal V}=\left(
\frac{3-2u^2}{u^2}\,\Delta-\frac{u^4-6u^2+6}{u^2\Delta}
\right)\frac{r^2N}{4}
-\left(
\frac{\Delta}{u^2}+\frac{(3-2u^2)(\beta^2-1)}{u^2\Delta}
\right)\frac{r^2}{4N}\,.
\ee
Inserting this to ${\cal H}$ in \eqref{HamA},
one can see that the equations 
$\partial{\cal H}/\partial N=0$ and 
$\partial{\cal H}/\partial \beta=0$
admit non-trivial solutions for $N,\beta$,
so that no constraints arise. 

Next, let us calculate 
the square root of the matrix \eqref{gamB}.
It can be chosen  in the form 
\be                                  \label{gamma}
\bs{\gamma}^\mu_{~\nu}=
\sqrt{g^{\mu\sigma}f_{\sigma\nu}}=\left(
\begin{array}{cccc}
a & {c} & 0 & 0 \\
-c & b & 0 & 0 \\
0 & 0 & u & 0 \\
0 & 0 & 0 & u
\end{array}
\right)\,,
\ee
and the conditions $\bs{\gamma}^\mu_{~\sigma}
\bs{\gamma}^\sigma_{~\nu}=
g^{\mu\sigma}f_{\sigma\nu}$
then reduce to 
\bea                    \label{abc}
a^2-c^2&=&\frac{1}{N^2}\equiv {\rm A},\nonumber \\
b^2-c^2&=&\Delta^2-\frac{\beta^2}{N^2}\equiv {\rm B},\nonumber \\
c\,(a+b)&=&\frac{\beta}{N^2}\equiv {\rm C},
\eea
and also $u^2={r^2}/{R^2}$. 
These equations can be rewritten as 
\be
a+b={\rm Y},~~~~~
a-b=\frac{{\rm A}-{\rm B}}{\rm Y},~~~~~c=\frac{\rm C}{\rm Y}\,,
\ee
where Y (not to be confused with $Y$ from \eqref{Y}) fulfills 
\be
{\rm Y}^4-2({\rm A}+{\rm B}){\rm Y}^2
+({\rm A}-{\rm B})^2-4{\rm C}^2=0.
\ee
Denoting Q$=\Delta/N$, this equation is solved by 
\be 
{\rm Y}=\sqrt{{\rm A}+{\rm B}+2{\rm Q}}=
\frac{1}{N}\sqrt{(N\Delta+1)^2-\beta^2 }\,,
\ee
so that 
\be
a=\frac{{\rm A}+{\rm Q}}{\rm Y},~~~~~
b=\frac{{\rm B}+{\rm Q}}{\rm Y},~~~~~
c=\frac{{\rm C}}{\rm Y},~~~~~
u=\frac{r}{R}\,.
\ee
Eigenvalues of 
$\bs{\gamma}^\mu_{~\nu}$
are 
\be                              \label{lambda}
\lambda_{0,1}=\left.\left.\frac12\right(a+b\pm\sqrt{(a-b)^2-4c^2}\right),
~~~\lambda_2=\lambda_3=u,
\ee
inserting which  into (\ref{4}) gives 
\bea                 \label{UUU}
{\cal U}_1&=&a+b+2u={\rm Y}+2u,\nonumber \\
{\cal U}_2&=&u(u+2a+2b)+ab+c^2=u(u+2{\rm Y})+{\rm Q}\,, \nonumber \\
{\cal U}_3&=&u\,(au+bu+2ab+2c^2)=u(u{\rm Y}+2{\rm Q}),\nonumber \\
{\cal U}_4&=&u^2(ab+c^2)=u^2{\rm Q}. 
\eea
As a result, 
the potential 
${\cal U}$ in (\ref{2}) is 
\be                   \label{UUU1}
{\cal U}=\sum_{k=0}^4 b_k {\cal U}_k=
P_0+P_1\,{\rm Y}
+P_2\,\frac{\Delta}{N}
\ee
with 
$P_m=b_{m}+2b_{m+1} u+b_{m+2} u^2$ for 
$m=0,1,2$. 
Multiplying  by $\sqrt{\h}N=NR^2/\Delta$ yields 
\be                    \label{V_dRGT}
{\cal V}=\frac{NR^2}{\Delta}\,P_0
+\frac{R^2P_1}{\Delta}\,\sqrt{(\Delta N+1)^2-\beta^2}
+R^2P_2\,,
\ee
which gives Eq.\eqref{XPEH} in the main text. 

\appendix
\setcounter{section}{2}
\setcounter{equation}{0}
\setcounter{subsection}{0}
\section{Fierz-Pauli limit  \label{ApC}}

\renewcommand{\theequation}{\Alph{section}.\arabic{equation}}

Eqs.\eqref{HamA},\eqref{H0rA},\eqref{V_dRGT} determine the  
Hamiltonian ${\cal H}=
N{\cal H}_0+\beta{\cal H}_r+m^2{\cal V}$ 
of the massive gravity theory. Let us consider its weak field limit, 
where 
\bea
\Delta=1+\delta ,~~~~~~
R=r+\rho\, ,~~~~~~
N=1+\nuu ,
\eea
with small $\delta$, $\rho$, $\nuu$ and where 
$\beta$, $\pii_\Delta$, 
$\pii_R$ are also small. 
One has 
\bea                \label{C2} 
{\cal H}_0&=&
\frac{\pii_\Delta^2}{4r^2}
+\frac{\pii_\Delta \pii_R}{2r}
+V_1+V_2+\ldots,~~~~
{\cal H}_r=\pii_\Delta^\prime+\pii_R+\ldots, \nonumber \\
{\cal V}&=&\nuu (r^2\delta -2r\rho)
+2r\delta\,\rho-\rho^2
-\frac{r^2}{4}\beta^2+\ldots\,,
\eea
where the dots denote higher order terms, while 
\bea
V_1&=&(2r(\delta+\rho^\prime))^\prime\,,~~~~~
V_2=2\rho\,\delta^\prime-\rho^{\prime 2}-\delta^2
+\mbox{total derivative}.
\eea
It is worth noting that 
both for the generic potential ${\cal V}$ in \eqref{V}
and for the dRGT potential \eqref{V_dRGT} 
the quadratic terms 
in the expansion in \eqref{C2} are the same.

Dropping the total derivatives and keeping only the 
quadratic terms,  the Hamiltonian density 
${\cal H}=
N{\cal H}_0+\beta{\cal H}_r+m^2{\cal V}$  reduces to 
\bea
{\cal H}_{\rm FP}&=&
\frac{\pii_\Delta^2}{4r^2}+\frac{\pii_\Delta \pii_R}{2r}
+V_2 
+\nuu\,V_1
+\beta(\pii_\Delta^\prime+\pii_R) \nonumber \\
&+&m^2\left(\nuu (r^2\delta -2r\rho)
+2r\delta\,\rho-\rho^2
-\frac{r^2}{4}\beta^2\right).
\eea
Varying this with respect to $\nu$ gives the constraint, 
\be                           \label{ApC5}
{\cal C}_{\rm FP}\equiv 
\frac{\partial{\cal H}_{\rm FP}}{\partial \nu}=
(2r(\delta+\rho^\prime))^\prime+
m^2(r^2\delta -2r\rho)=0,
\ee
while varying with respect to $\beta$ gives the 
equation 
\be 
\pii_\Delta^\prime+\pii_R-\frac{m^2r^2}{2}\,\beta=0\,,
\ee
so that 
\be 
\beta=\frac{2(\pii_\Delta^\prime +\pii_R )}{m^2 r^2}\,.
\ee
 Injecting this into ${\cal H}_{\rm FP}$, the result is
\be                           \label{ApC8}
{\cal H}_{\rm FP}={\cal E}_{\rm FP}+\nuu\,{\cal C}_{\rm FP}\,,
\ee 
where 
\be                           \label{Hfp1A}
{\cal E}_{\rm FP}(\pi)= 
\frac{\pii_\Delta^2}{4r^2}+\frac{\pii_\Delta \pii_R}{2r}
+\frac{(\pii_\Delta^\prime +\pii_R )^2}{m^2 r^2}+
2\rho\,\delta^\prime-\rho^{\prime 2}-\delta^2
+m^2(
2r\delta\rho-\rho^2).
\ee
Commuting ${\cal C}_{\rm FP}$ with the Hamiltonian 
$H_{\rm FP}=\int_0^\infty {\cal H}_{\rm FP} dr$
gives the second constraint,
\be 
{\cal S}_{\rm FP}=\{{\cal C}_{\rm FP},H_{\rm FP}\}=
\frac{m^2}{2}\,(r\pii_R-\pii_\Delta)
-(\pii_\Delta^{\prime}+\pii_R)^\prime=0\,.
\ee
These expressions for 
${\cal C}_{\rm PF}$, 
${\cal H}_{\rm PF}$,
${\cal S}_{\rm PF}$  
give rise to Eqs.\eqref{Hfp1B}--\eqref{S_FP} for the FP energy and constraints 
in the main text.
The same expressions 
can also be obtained
by inserting 
 the linearised 
\be 
\pi^i_k={\rm diag}\left[-\frac{\pii_\Delta}{2},
\frac{\pii_R}{4}, \frac{\pii_R}{4}\right],~~~~
\hh^i_k={\rm diag}\left[-2\delta,
\frac{2\rho}{r}, \frac{2\rho}{r}\right],~~~~
\ee
into Eqs.\eqref{CFP}, 
 \eqref{Hfp1}, \eqref{SFP}
with $f_{ik}dx^i dx^k=dr^2+r^2d\Omega^2$. 

It is instructive to derive Eqs.\eqref{ApC5},\eqref{ApC8},\eqref{Hfp1A}
once again 
via expanding the expressions \eqref{CON_C},\eqref{CON_S},\eqref{En} for the  
energy ${\cal E}_0$ and constraints ${\cal C,S}$ 
obtained in Section \ref{Sec4}. 
Expanding the 
constraints \eqref{CON_C},\eqref{CON_S} gives   
\be                   \label{C-expand}
0={\cal C}={\cal C}^{(1)}+{\cal C}^{(2)}
+\mbox{cubic and higher order terms},
\ee
with ${\cal C}^{(1)}={\cal C}_{\rm FP}$ and 
\bea 
{\cal C}^{(2)}={\cal E}_{\rm FP}
-\frac{(\pii_\Delta^\prime +\pii_R )^2}{2m^2 r^2}
+2\left((\rho+r\delta)\rho^\prime \right)^\prime 
+m^2[(c_3-2)\rho(\rho-2r\delta)-r^2\delta^2]
),
\eea
and also 
\be                   \label{S-expand}
0={\cal S}={\cal S}_{\rm FP}+
\mbox{quadratic and higher order terms}. 
\ee
One can see that the linear terms in the expansions 
\eqref{C-expand},\eqref{S-expand} agree 
with \eqref{C_FP},\eqref{S_FP}. 
Let us now expand the energy density \eqref{En}.
Dropping the total derivative, 
$${\cal E}_{0}=
{\cal E}^{(1)}_{0}+{\cal E}^{(2)}_{0}
+\mbox{cubic and higher order terms},
$$
with 
${\cal E}^{(1)}_{0}=-2\,{\cal C}_{\rm FP}$
and 
\bea
{\cal E}^{(2)}_{0}&=&
-\frac{\pii_\Delta^2}{4r^2}-\frac{\pii_\Delta \pii_R}{2r}
+\rho^{\prime 2}-2\rho\delta^\prime -2r\rho^\prime\delta^\prime 
+2\delta^2 \\
&+&m^2[(5-2c_3)\rho^2+4\,(c_3-3)\,r\rho\,\delta+3r^2\delta^2].
\nonumber 
\eea
Comparing with \eqref{Hfp1A}, 
one can see that 
${\cal E}_{0}$ looks actually quite different 
from ${\cal E}_{\rm FP}$,  so that 
one may wonder how the two expressions could 
agree with each other. 
 However, they completely agree when the 
constraints are imposed {up to the second order terms}. Indeed, 
according to Eq.\eqref{Hmm0} one has 
${\cal H}={\cal E}_{0}+(N+1/\Delta)\,{\cal C}$. 
Expanding this around flat space and comparing with \eqref{ApC8} 
gives the relation 
\be                                \label{EEE}
{\cal E}_{\rm FP}=
{\cal E}^{(1)}_{0}+{\cal E}^{(2)}_{0}
+2\,({\cal C}^{(1)}+{\cal C}^{(2)})
-\delta\,{\cal C}^{(1)}\,,
\ee
which can be directly verified. 
The constraint ${\cal C}=0$ implies that 
${\cal C}^{(1)}+{\cal C}^{(2)}=0$, up to higher order terms,
hence ${\cal C}^{(1)}=-{\cal C}^{(2)}$, and 
therefore 
the term $\delta\, {\cal C}^{(1)}$ is actually cubic 
in fields. As a result, the last three terms on the right in 
\eqref{EEE} do not contribute in the quadratic 
approximation, so that 
 ${\cal E}_{\rm FP}=
{\cal E}^{(1)}_{0}+{\cal E}^{(2)}_{0}={\cal E}_0$.

\appendix
\setcounter{section}{3}
\setcounter{equation}{0}
\setcounter{subsection}{0}
\section{Positivity of the Fierz-Pauli energy  \label{ApFP}}
\renewcommand{\theequation}{\Alph{section}.\arabic{equation}}
It is instructive to verify that the FP energy is indeed positive, which is 
not immediately obvious. 
The FP energy is 
\be                                \label{E_FPapp}
E_{\rm FP}=\int {\cal H}_{\rm FP}(\pi_{ik})\,d^3x
+\int {\cal H}_{\rm FP}(\hh_{ik})\,d^3x\,
\ee
with ${\cal H}_{\rm FP}(\pi_{ik})$ and ${\cal H}_{\rm FP}(\hh_{ik})$ defined 
by Eqs.\eqref{E-FP} and \eqref{Hfp1} in the main text,
where $h_{ik}$ and $\pi_{ik}$ 
should fulfill the constraints
\eqref{CFP} and \eqref{SFP}. Let us consider the Fourrier expansion,
\be 
\pi_{ik}({\bf x})=\frac{1}{(2\pi)^{3/2}}\int \Pi_{ik}({\bf k})e^{i{\bf kx}} d^3k\,,~~~
\ee 
with 
$ \Pi_{ik}({\bf k})=\Pi_{ik}^\ast(-{\bf k})$, the star denoting  
complex conjugation.  One has 
\be 
\int {\cal H}_{\rm FP}(\pi_{ik})\,d^3x
=\frac{1}{(2\pi)^{3/2}}\int {\cal E}({\bf k})d^3k \,,
\ee
where 
\be                         \label{ap1}
{\cal E}({\bf k})=2\sum_{i,k}|\Pi_{ik}|^2-
|\sum_s\Pi_{ss}|^2
+\frac{4}{m^2}\sum_i \left|\sum_s \Pi_{is}k^s\right|^2,
\ee
with $\Pi_{ik}\equiv\Pi_{ik}({\bf k})$. 
The constraint \eqref{SFP} requires 
\be                         \label{ap2}
m^2\sum_s\Pi_{ss}-2\sum_{i,s}k^i k^s \Pi_{is}=0.
\ee
The symmetric tensor $\Pi_{ik}({\bf k})$ can be expanded in the tensor basis,
\be                                \label{Papp}
\Pi_{ik}({\bf k})=\sum_{a=1}^6\phi_a({\bf k})\Pi^{(a)}_{ik},
\ee
where the tensors $\Pi^{(a)}_{ik}$
can be chosen to describe
two spin-2 tensor harmonics, two spin-1 vector harmonics,
and two scalar modes. Aligning the third coordinate axis along vector ${\bf k}$,
the tensor modes are traceless and orthogonal to ${\bf k}$, 
\be
\Pi^{(1)}_{ik}=
\frac{1}{\sqrt{2}}
\begin{pmatrix}
1 & 0 & 0 \\
0 & -1 & 0 \\
0 & 0 & 0 \\
\end{pmatrix},~~~~
\Pi^{(2)}_{ik}=
\frac{1}{\sqrt{2}}
\begin{pmatrix}
0 & 1 & 0 \\
1 & 0 & 0 \\
0 & 0 & 0 \\
\end{pmatrix},
\ee
the vector modes are 
\be
\Pi^{(3)}_{ik}=
\frac{1}{\sqrt{2}}
\begin{pmatrix}
0 & 0 & 1 \\
0 & 0 & 0 \\
1 & 0 & 0 \\
\end{pmatrix},~~~~
\Pi^{(4)}_{ik}=
\frac{1}{\sqrt{2}}
\begin{pmatrix}
0 & 0 & 0 \\
0 & 0 & 1 \\
0 & 1 & 0 \\
\end{pmatrix},
\ee
while the scalar modes 
\be 
\Pi^{(5)}_{ik}=\frac{1}{\sqrt{6}}\,{\rm diag}[1,1,-2],~~~~
\Pi^{(6)}_{ik}=\frac{1}{\sqrt{3}}\,\delta_{ik}\,,
\ee
so that
\be 
\sum_{ik}\Pi^{(a)}_{ik}\Pi^{(b)}_{ki}=\delta_{ab}.
\ee
Inserting \eqref{Papp} to \eqref{ap1}, ${\cal E}({\bf k})$ becomes 
\be
{\cal E}({\bf k})=2\sum_{a=1}^5 |\phi_a|^2-|\phi_6|^2
+2s^2(|\phi_3|^2+|\phi_4|^2) 
+\left.\left.\frac{4s^2}{3}\right|\sqrt{2}\,\phi_5-\phi_6\right|^2\,,
\ee
with $s^2={\bf k}^2/m^2$. This expression is not 
positive definite. 
However, the constraint \eqref{ap2} imposes the relation between the two 
scalar modes, 
\be 
\frac{2\sqrt{2}\,s^2}{\sqrt{3}}\,\phi_5+\sqrt{3}\left(
1-\frac{2s^2}{{3}}
\right)\phi_6=0,
\ee
which removes one of the two scalars. In view  of this 
the energy becomes 
\be 
{\cal E}({\bf k})=2(|\phi_1|^2+|\phi_2|^2)+
2(1+s^2)(|\phi_3|^2+|\phi_4|^2)
+\frac{9}{4s^4}|\phi_6|^2\,,
\ee
therefore $\int {\cal H}_{\rm FP}(\pi_{ik})\,d^3x\geq 0$.
Similarly one shows that $\int {\cal H}_{\rm FP}(h_{ik})\,d^3x\geq 0$.

\appendix
\setcounter{section}{4}
\setcounter{equation}{0}
\setcounter{subsection}{0}
\section{Poisson brackets  \label{ApE}}
\renewcommand{\theequation}{\Alph{section}.\arabic{equation}}

Some care is needed when computing the Poisson brackets
(see, for example, \cite{Khoury:2011ay}).
Let us denote by $q_k=(\Delta,R)$ and $p_k=(p_\Delta,p_R)$ 
the phase space coordinates and their momenta, they depend on time 
and on the radial coordinate $r$.
Consider a function on the phase space, 
\be
{\cal F}=(q_k,q_k^\prime,q_k^{\prime\prime},\ldots q_k^{(M)},
p_k,p_k^\prime,p_k^{\prime\prime},\ldots p_k^{(M)},r)\,,
\ee 
where the primes denote derivatives with respect to $r$, while  $M$ is the 
order of the highest 
derivative. One considers the functional 
\be
F=\int_0^\infty f(r){\cal F}\, dr\,,
\ee
where $f$ is a smoothening function, which is assumed to vanish fast enough 
for $r\to 0,\infty$ in order that one could integrate by parts and always
drop the boundary terms. The variation of $F$ is 
\be
\delta F=\int  f(r)\sum_k\left(
\frac{\partial{\cal F}}{\partial q_k}\,\delta q_k 
+\ldots 
+\frac{\partial{\cal F}}{\partial q^{(M)}_k}\,\delta q^{(M)}_k 
+
\frac{\partial{\cal F}}{\partial p_k}\,\delta p_k 
+\ldots 
+\frac{\partial{\cal F}}{\partial p^{(M)}_k}\,\delta p^{(M)}_k 
\right)dr
\ee
and integrating by parts,
\bea
\delta F=\int dr \sum_k\left(
f\frac{\partial{\cal F}}{\partial q_k}
-\left(f\frac{\partial{\cal F}}{\partial q^\prime_k}\right)^\prime
+\ldots 
+(-1)^M\left(f\frac{\partial{\cal F}}{\partial q^{(M)}_k}\right)^{(M)}\right)\delta q_k 
\nonumber \\
+
\int dr \sum_k\left(
f\frac{\partial{\cal F}}{\partial p_k}
-\left(f\frac{\partial{\cal F}}{\partial p^\prime_k}\right)^\prime
+\ldots 
+(-1)^M\left(f\frac{\partial{\cal F}}{\partial p^{(M)}_k}\right)^{(M)}\right)\delta p_k \,.
\eea
Therefore, the functional derivatives are 
\bea                            \label{dF}
\frac{\delta F}{\delta q_k}=
f\frac{\partial{\cal F}}{\partial q_k}
-\left(f\frac{\partial{\cal F}}{\partial q^\prime_k}\right)^\prime
+\ldots 
+(-1)^M\left(f\frac{\partial{\cal F}}{\partial q^{(M)}_k}\right)^{(M)}\,,\nonumber \\
\frac{\delta F}{\delta p_k}=
f\frac{\partial{\cal F}}{\partial p_k}
-\left(f\frac{\partial{\cal F}}{\partial p^\prime_k}\right)^\prime
+\ldots 
+(-1)^M\left(f\frac{\partial{\cal F}}{\partial p^{(M)}_k}\right)^{(M)}\,.
\eea
Let us consider another function on the phase space, 
\be
{\cal G}(q_k,q_k^\prime,q_k^{\prime\prime},\ldots q_k^{(M)},
p_k,p_k^\prime,p_k^{\prime\prime},\ldots p_k^{(M)},r)\,,
\ee 
whose smoothened version is 
\be
G=\int_0^\infty g(r){\cal G}\, dr\,,
\ee
with another smoothening function $g(r)$. The functional derivatives are 
\bea                       \label{dG}
\frac{\delta G}{\delta q_k}=
g\frac{\partial{\cal G}}{\partial q_k}
-\left(g\frac{\partial{\cal G}}{\partial q^\prime_k}\right)^\prime
+\ldots 
+(-1)^M\left(g\frac{\partial{\cal G}}{\partial q^{(M)}_k}\right)^{(M)}\,,\nonumber \\
\frac{\delta G}{\delta p_k}=
g\frac{\partial{\cal G}}{\partial p_k}
-\left(g\frac{\partial{\cal G}}{\partial p^\prime_k}\right)^\prime
+\ldots 
+(-1)^M\left(g\frac{\partial{\cal G}}{\partial p^{(M)}_k}\right)^{(M)}\,.
\eea
The Poisson bracket is defined as 
\bea                                             \label{BP}
\{F,G\}_{\rm PB}&=&\int_0^\infty dr \sum_k\left(
\frac{\delta F}{\delta q_k}\frac{\delta G}{\delta p_k}
-
\frac{\delta F}{\delta p_k}\frac{\delta G}{\delta q_k}
\right) \nonumber \\
&\equiv& 
\int_0^\infty\int_0^\infty dr\, ds\, f(r)g(s)\,\{{\cal F}(r),{\cal G}(s)\}_{\rm PB}. 
\eea
To compute the integrand 
in the first line here  one uses 
 the definitions \eqref{dF},\eqref{dG}, while the passage  
to the second line is achieved by inserting the 
delta-functions. For the analysis in the main body of the paper 
it is sufficient to calculate only the first line in \eqref{BP}. This only requires 
implementing the definitions \eqref{dF},\eqref{dG}, which can be efficiently 
done with MAPLE, say. 
For example, for ${\cal H}_0$ and ${\cal H}_r$ defined in \eqref{H0r} one 
obtains 
\bea
\{H_0,H_0\}_{\rm PB}&=&\int_0^\infty dr (fg^\prime-f^\prime g)\,\Delta^2 {\cal H}_r\,,~\nonumber \\
\{H_0,H_r\}_{\rm PB}&=&\int_0^\infty dr f\left(g{\cal H}_0\right)^\prime\,,\nonumber \\
\{H_r,H_r\}_{\rm PB}&=&\int_0^\infty dr (fg^\prime-f^\prime g) {\cal H}_r\,.
\eea
Inserting here the delta functions gives the $t,r$ part of the 
diffeomorphism algebra \cite{Khoury:2011ay},
\bea
\{{\cal H}_0(r),{\cal H}_0(s)\}_{\rm PB}
&=&{\cal H }^r(r)\partial_r\,\delta(r-s)-{\cal H }^r(s)\partial_s\,\delta(r-s),  \nonumber \\
\{{\cal H}_0(r),{\cal H}_r(s)\}_{\rm PB}
&=&{\cal H }_0(s)\partial_r\,\delta(r-s), \nonumber \\
\{{\cal H}_r(r),{\cal H}_r(s)\}_{\rm PB}
&=&{\cal H }_r(r)\partial_r\,\delta(r-s)-{\cal H }_r(s)\partial_s\,\delta(r-s) .
\eea
Next, when computing the commutator of the constraint ${\cal C}(r)$ with itself
one obtains 
\be
\{C,C\}_{\rm PB}=\int_0^\infty dr (fg^\prime-f^\prime g)\,\frac{\Delta^2}{Y}\, {\cal C}\,,\nonumber \\ 
\ee
with $Y$ from \eqref{Y}, 
from where it follows that  $\{{\cal C}(r_1),{\cal C}(r_2)\}_{\rm PB}=0$ if ${\cal C}=0$. 
Similarly, the secondary constraint ${\cal S}(r)=\{{\cal C}(r),H\}$ is obtained by computing
$\{C,H\}_{\rm PB}$, which yields and expression of the form 
\be
\int_0^\infty dr (fA_0+f^\prime A_1+\ldots +A_M f^{(M)})\,.\nonumber \\ 
\ee
Integrating by parts brings this to 
\be
\int_0^\infty dr \left(A_0-(A_1)^\prime+\ldots +(-1)^M(A_M)^{(M)}\right) f\,,
\ee
and setting $f=g=1$ the integrand gives ${\cal S}$.


%

\end{document}